\documentclass[paper,notoc,nohyper]{JHEP}
\usepackage{graphics}
\newcommand{\be}{\begin{equation}}
\newcommand{\ee}{\end{equation}}
\newcommand{\bea}{\begin{eqnarray}}
\newcommand{\eea}{\end{eqnarray}}

\def\Lx{X}
\def\Ly{Y}
\def\Libx{{\rm Li}_2(x)}
\def\Liby{{\rm Li}_2(y)}
\def\Licx{{\rm Li}_3(x)}
\def\Licy{{\rm Li}_3(y)}

\def\Lidx{{\rm Li}_4(x)}
\def\Lidy{{\rm Li}_4(y)}
\def\Lidz{{\rm Li}_4(z)}
\def\bb{}
\def\PR{\left(\frac{1+\gamma_5}{2}\right)}
\def\PL{\left(\frac{1-\gamma_5}{2}\right)}
\def\sss#1{\slash \!\! #1}
\def\bb{}

\def\sstt{\frac{t}{u}}

\def\utss{ }

\def\tttu{\frac{t^2}{s^2}}
\def\tfiveou{\frac{t^4}{u^2s^2}}

\def\M#1#2{{\cal M}^{({#1})}_{#2}}
\def\MF#1#2{{\cal M}^{({#1}),F}_{#2}}
\def\MP#1#2{{\cal M}^{({#1}),P}_{#2}}
\def\MS#1#2{{\cal M}^{({#1}),S}_{#2}}
\def\MV#1#2{{\cal M}^{({#1}),V}_{#2}}
\def\MX#1#2{{\cal M}^{({#1}),X}_{#2}}

\title{\boldmath Two-loop corrections to Light-by-Light scattering in 
Supersymmetric QED\footnote{Work supported in part by the UK Particle Physics and Astronomy 
Research Council, by the EU Fifth Framework Programme `Improving Human
Potential', Research Training Network `Particle Physics Phenomenology 
at High Energy Colliders', contract HPRN-CT-2000-00149 and
by the DFG-Forschergruppe
Quantenfeldtheorie, Computeralgebra und Monte-Carlo Simulation.
We thank the
British Council and German Academic Exchange Service for support under ARC
project 1050.
PM acknowledges the support of the German Academic Exchange Service.}}
\author{
T.~Binoth$^a$,
E.~W.~N.~Glover$^b$,
P.~Marquard$^c$ and J.~J.~van der Bij$^c$\\
$^a$Department of Physics and Astronomy,\\
           The University of Edinburgh,
	   EH9 3JZ Edinburgh, Scotland
	   \\[1mm]
	   $^b$Department of Physics, 
University of Durham, 
Durham DH1 3LE, 
England\\[1mm]
$^c$Fakult{\"a}t f{\"u}r Physik,
        Albert-Ludwigs-Universit{\"a}t Freiburg,
        D-79104 Freiburg, Germany\\[1mm] 
E-mail: \email{binoth@ph.ed.ac.uk}, \email{E.W.N.Glover@durham.ac.uk},
\email{marquard@physik.uni-freiburg.de}, 
\email{jochum@phyv1.physik.uni-freiburg.de}}
\abstract{We use an $n$-dimensional projection method to calculate
helicity amplitudes for photon-photon scattering up to the two-loop level,
for massless particles inside the loop.
We confirm the recent standard model results found by Bern et al using
another calculational method.
We further apply the method to $N=1$ and $N=2$ supersymmetric
QED. As predicted by the supersymmetric Ward identity, two of the independent
amplitudes vanish. The remaining amplitude simplifies as the number of
supersymmetries increases.   For $N=2$ SUSY QED, only contributions 
of weight 4 remain.
 }

\preprint{{DTP/02/34}, {IPPP/02/17}, {EDINBURGH--2002--02},
{Freiburg-THEP-02/04}, {hep-ph/0202266}}
\begin{document}

\section{Introduction}

Light-by-light scattering is a paradigm process in quantum field theory even if
it is not yet of experimental relevance.\footnote{See the nice review of the
current experimental situation in Ref.~\cite{Bern:2001dg}.}   The calculation
of the process at the one-loop level, the first non vanishing order, was
accomplished quite some time
ago~\cite{Karplus:1950,Karplus:1951,DeTollis:1964,DeTollis:1965,Costantini:1971}.
Gauge invariance and IR/UV finiteness leads to enormous cancellation mechanisms
when individual Feynman graphs are summed up. As such the study of
light-by-light scattering is an ideal testing ground for new methods relevant
for loop calculations.

The computation of the higher order corrections to this process  has been
plagued by the  difficulty of evaluating two-loop four point functions with
four on-shell legs. At present 
it is not possible to include the mass of the
particles circulating inside the loops. However, the complete set of integrals
for planar on-shell  two-loop graphs with massless internal particles is now
known~\cite{Smirnov:1999gc,Smirnov:1999wz,Tausk:1999vh,Anastasiou:2000mf,Anastasiou:1999bn,
Gehrmann:2000xj, Anastasiou:2000kp} together with a set of algorithms for
reducing the tensor integrals down to the basis set of master
integrals~\cite{Chetyrkin:1980pr,Chetyrkin:1981qh,Gehrmann:2000as}.   This
technology has recently been applied to calculate the two-loop matrix elements
for a wide range of $2\to 2$ scattering
processes~\cite{Bern:2000dn,Bern:2001ie,Anastasiou:2001kg,Anastasiou:2001ue,Anastasiou:2001sv,Glover:2001af,Bern:2001df,Bern:2001dg,Bern:2002tk,Anastasiou:2002zn}.   

Recently the two-loop corrections for light-by-light scattering were
calculated~\cite{Bern:2001dg} by applying the helicity formalism at the
two-loop level. 
The helicity formalism, which is strictly defined only in
4-dimensions greatly simplifies the calculation on the
one-loop level. The generalisation to the two-loop case
was done only recently~\cite{Bern:2000dn,Bern:2001df,Bern:2001dg,Bern:2002tk}.
In other cases~\cite{Bern:2001ie,Anastasiou:2001kg,
Anastasiou:2001ue,Anastasiou:2001sv,Glover:2001af,Anastasiou:2002zn} the interference between
two-loop and
tree level amplitudes directly relevant for next-to-next-to-leading order
calculations  has been computed by working with $n$-dimensional kinematics only.  This
approach is not applicable in the case of four-photon scattering as a tree
level operator simply does not exist.

In the present paper we follow another approach to calculate helicity
amplitudes which in a certain sense interpolates between both methods.
Namely by analysing the tensorial structure of the
amplitude~\cite{Karplus:1950}, valid to all orders in perturbation theory,
we define projectors to isolate the coefficients of particular tensor
structures.  These projectors are defined in $n$-dimensions and therefore
allow for a completely $n$-dimensional treatment of the computation of the
tensor coefficients. The symmetry of the process dictates that there are
only three independent  coefficients.  By fixing the helicity of the
external photons,  it is then straightforward to apply helicity methods to
extract specific helicity amplitudes from the general tensor in terms of
the three $n$-dimensional coefficients.  The relation between the helicity
amplitude and the tensor coefficients is independent of the order that the
coefficients are computed at. It is also independent of whether or not the
particles circulating in the loops are massive. As an explicit
example, we focus on the circularly polarised amplitudes. Similar
techniques can be applied to obtain linear polarised amplitudes.  Note
that  in defining the helicity amplitudes the external states are
four-dimensional, however, the tensor coefficients are fully
$n$-dimensional.

We apply the method to a fairly general class of Lagrangians containing
photons, charged and neutral scalars and charged and neutral fermions with
prescribed couplings. There are several gauge invariant classes  and we list
these contributions separately. We give explicit results for the $N=1$ and
$N=2$ SUSY QED Lagrangians~\cite{Wess:1974jb,Hollik:1999xh,Fayet:1976yi}.
Although these  unbroken supersymmetric theories have no direct
phenomenological relevance, they have a  theoretical interest of their own and
have led to new insights  in the possibilities of quantum field theory, in
particular  regarding its divergences. The presence of a supersymmetry leads to
important cancellations between the bosonic and fermionic degrees of freedom.
One of the best known consequences is the disappearance of quadratic
divergences in $N=1$ supersymmetric theories. These divergences generally
appear in renormalization constants and have no direct physical consequences.
More interesting are cancellations or simplifications for physical quantities.
These can be analysed on the basis of supersymmetric Ward
identities~\cite{Grisaru:1977vm,Grisaru:1977px,Parke:1985pn}. In the case at
hand these lead to the prediction that two of the independent helicity
amplitudes vanish, a fact that we will verify in our calculation. The general
analysis of supersymmetric Ward identities is a rather involved affair, due to
the fact that there is no easily implementable supersymmetric regularization.
The identification of a supersymmetry preserving regulator is beyond the scope
of this paper. We therefore work in conventional dimensional regularisation
which explicitly breaks supersymmetry by keeping $n-2$ polarizations for
internal photons while the photino has only 2 degrees of freedom.  However,
because  the quantities we calculate  are infrared and ultraviolet finite, they
are supersymmetric in the four-dimensional limit. There may be violations of
supersymmetry of ${\cal O}(n-4)$, but they are not relevant in the
four-dimensional limit. 

While the simplifications in $N=1$ supersymmetry are already very interesting,
stronger results are possible in the case of $N=2$ supersymmetry. For instance
it is known that the beta function in $N=2$ super Yang-Mills theories is fully
determined at the one-loop level.  Since the infinite contributions that
determine the beta function satisfy such strong non-renormalization theorems,
one can also expect large cancellations for finite quantities.  In order to
check for such cancellations in the photon-photon scattering amplitude it is
necessary to work at the two-loop level, since at the one-loop level there is 
no difference between $N=1$ and $N=2$ supersymmetry. We will indeed find  an
interesting pattern of simplifications taking place going up in the number of
supersymmetries. Similar simplifications
for certain supersymmetric higher loop amplitudes
were already reported in the case of $N=4$ Super-Yang-Mills \cite{Bern:1997nh}
and $N=8$ Supergravity \cite{Bern:1998ug}.

Our paper is organised as follows.   In Sec.~\ref{sec:model} we recall the
Lagrangian for Supersymmetric QED and review the particle content of the
theory. The relevant Feynman rules are collected in Appendix~\ref{sec:rules}.
The general tensor structure for light-by-light scattering is detailed in
Sec.~\ref{sec:tensor}, together with the constraints from transversality of the
external polarisation vectors, Bose symmetry and the gauge Ward identities that
reduce the number of independent tensor coefficients to three.   The general
tensor is mapped onto circularly polarised helicity amplitudes in
Sec.~\ref{subsec:helamp}. We explicitly show how each of the three independent
helicity amplitudes can be written in terms of the three $n$-dimensional tensor
coefficients.  In Sec.~\ref{subsec:projectors} we introduce projection
operators that can be applied to the full tensor to isolate any of the tensor
coefficients.  A discussion of the Supersymmetric Ward identities is given in
Sec.~\ref{subsec:SWI} while the results of an explicit calculation of the one-
and two-loop helicity amplitudes are given in Secs.~\ref{subsec:oneloop} and
\ref{subsec:twoloop} respectively.  The individual contributions to the
two-loop amplitudes from the gauge invariant sub-classes of diagrams are listed
in Appendix~\ref{sec:conts}. Finally our findings are summarised in
Sec.~\ref{sec:summary}.

\section{The SUSY QED Lagrangian}
\label{sec:model}

$N=1$ Supersymmetric QED~\cite{Wess:1974jb,Hollik:1999xh} 
is an abelian gauge theory
containing a vector multiplet 
$(A^\mu,\lambda_\alpha,\bar\lambda^{\dot\alpha})$ containing the 
vector photon field and the Majorana photino and two chiral multiplets
$(\psi_L^\alpha,\phi_L^-)$ and $(\psi_R^\alpha,\phi_R^+)$ 
with charge $Q_L=-1$ and $Q_R=+1$, each consisting of a Weyl spinor and 
a scalar field representing  
the electron and scalar electron matter fields.
The $N=1$ SUSY QED Lagrangian with massless matter fields is given by,
\begin{eqnarray}
{\cal L}_{\rm SQED}^{N=1} & = &
-\frac{1}{4}F_{\mu\nu}F^{\mu\nu} +
\frac{1}{2}\overline{\tilde{\gamma}_1}i\gamma^\mu\partial_\mu\tilde{\gamma}_1
\nonumber\\
&&{}
+|D_\mu \phi_L^-|^2 + |D^\dagger_\mu \phi_R^+|^2
+ \overline{\Psi}i\gamma^\mu D_\mu \Psi
\nonumber\\
&&{}
+\sqrt{2}e\left(
\overline{\Psi}P_R\tilde{\gamma}_1\phi_L^-
- \phi_R^+ \overline{\tilde{\gamma}_1}P_R \Psi + h.c. \right)
\nonumber\\
&&{}
-\frac{1}{2}e^2 \left(|\phi_L^-|^2 -|\phi_R^+|^2\right)^2,
\label{LSQED1}
\end{eqnarray}
where
\begin{equation}
\Psi= {\psi_{L\alpha} \choose \overline{\psi}_R^{\dot\alpha}},\qquad 
\tilde\gamma_1 = {-i\lambda_\alpha \choose i\bar\lambda^{\dot\alpha}},
\end{equation}
and where $D_\mu$ is the gauge covariant derivative and $F_{\mu\nu}$ is the
field strength of the photon.   This Lagrangian corresponds to the
Wess-Zumino gauge where all unphysical fields are removed by gauge
transformations.

$N=2$ Supersymmetric QED~\cite{Wess:1974jb} contains in
addition a chiral multiplet,
a Majorana gaugino and a complex scalar photon, 
 $(\psi^0_\alpha, \bar \psi^{0\dot\alpha},\phi^0)$.   
The $N=2$ SUSY QED Lagrangian with massless matter fields is given by,
\begin{eqnarray}
{\cal L}_{\rm SQED}^{N=2} & = & {\cal L}_{\rm SQED}^{N=1} 
+ \frac{1}{2}\overline{\tilde{\gamma}_2}i\gamma^\mu\partial_\mu\tilde{\gamma}_2
+|\partial_\mu \phi^0|^2  
\nonumber\\
&&
+\sqrt{2}e\left(
\overline{\Psi}P_L\tilde{\gamma}_2\phi_L^-
+ \phi_R^+ \overline{\tilde{\gamma}_2}P_L \Psi + h.c. \right)
\nonumber\\
&&
+\sqrt{2}e\left(
\overline{\Psi}P_L\Psi\phi^0
+ \phi^{0*} \overline{\Psi}P_R \Psi  \right)
\nonumber\\
&&{}
-2e^2 \left(|\phi_L^-|^2 + |\phi_R^+|^2 \right)^2 | \phi^0|^2\nonumber \\
&&{}
-\frac{1}{2}e^2 \left(|\phi_L^-|^2 + |\phi_R^+|^2 \right)^2
+\frac{1}{2}e^2 \left(|\phi_L^-|^2 - |\phi_R^+|^2 \right)^2,
\label{LSQED2}
\end{eqnarray}
where
$$
\tilde\gamma_2 = \left(\begin{array}{c} \psi_{0\alpha} 
\\ \overline{\psi}_0^{\dot\alpha} \end{array}\right)
$$
Note that the $N=2$ supersymmetry leads to   
a $SU(2)$ symmetry under which the photinos and scalar
electrons transform nontrivially.
By writing them as doublets 
$$
\left(\begin{array}{cc}
\phi_L^+ \\ \phi_R^+
\end{array}\right) \; , \; 
\left(\begin{array}{cc}
\tilde\gamma_1 \\ \tilde \gamma_2
\end{array}\right),
$$
the photino--electron--selectron interactions can then be written as
\begin{eqnarray}
{\cal L}_{\gamma e\tilde e} =  
 \sqrt{2}e (\phi_L^-,\phi_R^-)~
\left[ \overline{\Psi} P_R 
\left(\begin{array}{cc}
1 & 0 \\ 0 & 1
\end{array}\right)
+ \overline{\Psi} P_L 
\left(\begin{array}{cc}
0 & 1 \\ -1 & 0
\end{array}\right) \right] 
\left(\begin{array}{c}
\tilde\gamma_1 \\ \tilde\gamma_2 
\end{array}\right) + h.c. 
\end{eqnarray}
which is equivalent to the manifestly SU(2)
invariant form presented in~\cite{Fayet:1976yi}.

For completeness, the Feynman rules for these Lagrangians are given in
Appendix~\ref{sec:rules}.

\section{The tensor structure of the four photon amplitude}
\label{sec:tensor}

We consider the process where all particles are incoming
\begin{equation}
\gamma(p_1, \lambda_1)
+\gamma(p_2, \lambda_2)
+\gamma(p_3, \lambda_3)
+\gamma(p_4, \lambda_4)\to 0,
\end{equation}
and photon $i$ carries momentum $p_i$ and has helicity $\lambda_i$.
The amplitude has the form
\bea
{\cal M} = 
\varepsilon_{1,\mu_1}\varepsilon_{2,\mu_2}\varepsilon_{3,\mu_3}\varepsilon_{4,\mu_4} 
{\cal M}^{\mu_1 \mu_2\mu_3\mu_4}(p_1,p_2,p_3,p_4),
\eea
where the scattering tensor ${\cal M}^{\mu_1 \mu_2\mu_3\mu_4}$ 
has the following general decomposition
\bea\label{tensordec}
{\cal M}^{\mu_1 \mu_2\mu_3\mu_4} &=& 
        A_1 g^{\mu_1\mu_2} g^{\mu_3\mu_4}
       +A_2 g^{\mu_1\mu_3} g^{\mu_2\mu_4}  
       +A_3 g^{\mu_1\mu_4} g^{\mu_2\mu_3} \nonumber \\
&&+\sum\limits_{j_1,j_2=1}^{3} 
\Bigl(
   B^1_{j_1 j_2}\,g^{\mu_1\mu_2}\,p_{j_1}^{\mu_3}\,p_{j_2}^{\mu_4}
  +B^2_{j_1 j_2}\,g^{\mu_1\mu_3}\,p_{j_1}^{\mu_2}\,p_{j_2}^{\mu_4}
  +B^3_{j_1 j_2}\,g^{\mu_1\mu_4}\,p_{j_1}^{\mu_2}\,p_{j_2}^{\mu_3}\nonumber \\ 
&&\phantom{\sum\limits_{j_1,j_2=1}^{3}}
+B^4_{j_1 j_2}\,g^{\mu_2\mu_3}\,p_{j_1}^{\mu_1}\,p_{j_2}^{\mu_4}
  +B^5_{j_1 j_2}\,g^{\mu_2\mu_4}\,p_{j_1}^{\mu_1}\,p_{j_2}^{\mu_3} 
  +B^6_{j_1 j_2}\,g^{\mu_3\mu_4}\,p_{j_1}^{\mu_1}\,p_{j_2}^{\mu_2}
\Bigr) 
\nonumber \\  
 &&+\sum\limits_{j_1,j_2,j_3,j_4=1}^{3} C_{j_1 j_2 j_3 j_4}
p_{j_1}^{\mu_1}\,p_{j_2}^{\mu_2} p_{j_3}^{\mu_3}\,p_{j_4}^{\mu_4}.
\eea
Terms such as
$\epsilon^{\mu_1\mu_2\mu_3\mu_4}$ are forbidden on the grounds of parity
invariance.
The functions $A_i$, $B^i_{jk}$ and $C_{ijkl}$ are functions of the 
Mandelstam variables, $s = (p_1+p_2)^2$, $t = (p_2+p_3)^2$, $u = (p_1+p_3)^2$
and the spacetime dimension $n$.  They also depend on the mass of the
electron and any other particles that may be involved in the scattering
process.
This decomposition is valid for arbitrary loop order since 
it is based solely on
the Lorentz structure of the external particles.
Altogether, there are 138 coefficients.
However, many coefficients are irrelevant since 
they drop out when contracted with the photon
polarisation vectors due to the transversality condition,
\be
\varepsilon_j\cdot p_j = 0,\quad \mbox{for } j \in \{1,2,3,4\}.
\ee
This reduces the number of coefficients to  57. 

Bose symmetry of the external photons means that these coefficients are not
independent.  Requiring invariance under the exchange of any index pair
$(j,\mu_j)$ where the first labels the external vector and the second is the
Lorentz index of the corresponding  polarisation vector reduces the number of
independent functions to 11.

The number of independent functions is further 
reduced by gauge symmetry. The related Ward identities
read in an obvious notation,
\be\label{ward_id}
{\cal M}^{p_1\varepsilon_2\varepsilon_3\varepsilon_4} =
{\cal M}^{\varepsilon_1 p_2 \varepsilon_3\varepsilon_4} =
{\cal M}^{\varepsilon_1\varepsilon_2 p_3\varepsilon_4} =
{\cal M}^{\varepsilon_1\varepsilon_2\varepsilon_3 p_4} = 0.
\ee 
Applying the gauge symmetry reduces the number of 
independent functions to three which we take to be,
\bea 
\label{eq:basis}
A_1(s,t,u), \qquad B^1_{11}(s,t,u), \qquad C_{2111}(s,t,u).
\eea
Once these functions are known, the full tensor can be reconstructed.

\subsection{Helicity amplitudes}
\label{subsec:helamp}

It is often convenient to express the amplitude in terms of the helicity
structure of the scattering process.   This can straightforwardly be
achieved using the Lorentz structure of the tensor. In principle there are
16 helicity amplitudes - two polarisations for each photon.   However,
parity, time-reversal and Bose symmetry reduce this number to four 
which is further reduced to three by crossing symmetry.   Note that the
number of independent helicity amplitudes matches the number of
independent functions describing the tensor. We take  
\be 
{\cal M}_{++++}, \qquad {\cal M}_{+++-},\qquad {\cal M}_{++--}, 
\ee 
to be a basis from which the other helicity amplitudes can be reconstructed.

By specifying polarisation vectors for the external photons and applying
them to the full tensor, we can derive the helicity amplitudes in terms of
the three independent functions $A_1$, $B_{11}^1$ and $C_{2111}$. We find
that, up to overall phases,
\begin{eqnarray}\label{ampPPPP}
{\cal M}_{++++}&=&A_1(s, t, u) + A_1(t, u, s) + A_1(u, t, s) \nonumber \\
&& - \frac{u^2}{4t}  B_{11}^1(s, t, u) 
- \frac{t  (2  s + t)}{4u}  B_{11}^1(s, u, t)  
- \frac{u^2}{4s}  B_{11}^1(t, s, u)   \nonumber\\
&&-  \frac{s  (t - u)}{4u}  B_{11}^1(t, u, s)  
+ \frac{(2  s - t)  t}{4s}  B_{11}^1(u, s, t)  
- \frac{s  (s - 2  t)}{4t} B_{11}^1(u, t, s) \nonumber\\
&& + \frac{s  (s - 2  t)  u}{8t}  C_{2111}(s, t, u)  
+ \frac{s  t  (t - u)}{8u}  C_{2111}(s, u, t)  
 + \frac{t^2  u }{8s} C_{2111}(t, s, u) \nonumber\\
 && - \frac{s  t^2}{8u}  C_{2111}(t, u, s)   
+ \frac{t  u^2}{8s}  C_{2111}(u, s, t)  
+ \frac{s  u^2}{8t} C_{2111}(u, t, s) ,
\end{eqnarray}
\begin{eqnarray}\label{ampPPPM}
{\cal M}_{+++-}&=&-\frac{u^2}{4t}  B_{11}^1(s, t, u)  
- \frac{t^2}{4u}  B_{11}^1(s, u, t)  
+ \frac{u^2}{4s}  B_{11}^1(t, s, u) \nonumber \\
 &&+ \frac{s  (t - u)}{4u} B_{11}^1(t, u, s)  
 + \frac{t^2}{4s}  B_{11}^1(u, s, t)  
 + \frac{s  (-t + u)}{4t}  B_{11}^1(u, t, s)  \nonumber\\
&&
+  \frac{s  (t - u)u}{8t}  C_{2111}(s, t, u) 
- \frac{s  t  (t - u)}{8u}  C_{2111}(s, u, t)  
-  \frac{t^2  u}{8s}  C_{2111}(t, s, u)  \nonumber\\
&&
+ \frac{s  t^2}{8u}  C_{2111}(t, u, s)  
- \frac{t  u^2}{8s}  C_{2111}(u, s, t)  
+ \frac{s  u^2}{8t}  C_{2111}(u, t, s) ,
\end{eqnarray}
and
\begin{eqnarray}\label{ampPPMM}
{\cal M}_{++--}&=& A_1(s, t, u) + A_1(t, u, s) + A_1(u, t, s) \nonumber \\
&&
- \frac{u^2}{4t}  B_{11}^1(s, t, u)  
- \frac{t  (2  s + t)}{4u}  B_{11}^1(s, u, t)  
- \frac{u^2}{4s}  B_{11}^1(t, s, u)  \nonumber\\
&&
- \frac{s  (t - u)}{4u}  B_{11}^1(t, u, s)  
+ \frac{(2  s - t)  t}{4s}  B_{11}^1(u, s, t)  
- \frac{s  (s - 2  t)}{4t}  B_{11}^1(u, t, s)  \nonumber\\
&&
 + \frac{s  (s - 2  t)  u}{8t}  C_{2111}(s, t, u)  
+ \frac{s  t  (t - u)}{8u}  C_{2111}(s, u, t)  
 + \frac{t^2  u}{8s}  C_{2111}(t, s, u)  \nonumber\\
&&
 - \frac{s  t^2}{8u}  C_{2111}(t, u, s)  
+ \frac{t  u^2}{8s}  C_{2111}(u, s, t)  
+  \frac{s  u^2}{8t}  C_{2111}(u, t, s) .
\end{eqnarray}
Amplitudes for linearly polarised light can also be straightforwardly obtained
from the general tensor and are given in terms of $A_1$, $B_{11}^1$ and
$C_{2111}$.

Note that in deriving the helicity amplitudes,  we have made no
assumptions about how the functions $A_1$, $B_{11}^1$ and $C_{2111}$ are
calculated. In the conventional approach to computing loop
helicity amplitudes one has to define a description to deal with
scalar products between loop momenta and polarisation vectors.
 Furthermore, we have made no assumptions about the 
masses of particles in the loops.

\subsection{Projection operators}
\label{subsec:projectors}

To calculate the independent functions $A_1$, $B_{11}^1$ and $C_{2111}$ it
is convenient to define projection operators that (a) isolate the function
and (b) saturate the free Lorentz indices. This latter point is important
for  practical calculations since it allows for the cancellation of
reducible scalar products between loop momenta  and external vectors in
the Feynman diagram integrals. These reducible scalar products  can be
expressed in terms of inverse propagators and many cancellations already
happen at this level. For example at the one-loop level, no tensor box 
integral needs to be evaluated since any loop momentum appearing in the
numerator  will immediately cancel one of the propagators. 
This reduces the complexity
of the one-loop calculation enormously. 
For higher-loop or multi-leg applications
such simplifications are highly appreciated. 

To invert Eq. (\ref{tensordec}) and isolate the independent functions,
it is useful to define the following tensors,
\bea
{\cal P}^{\mu\nu} &=& g^{\mu\nu} 
-  2 \sum_{j=1}^3 \sum_{k=1}^3 \, p_{j}^{\mu}\, H_{jk}\, p_{k}^{\nu}, \\
{\cal R}_j^{\nu}  &=& 2 \sum_{k=1}^3 \,H_{jk}\, p_{k}^{\nu},
\eea
where $H=G^{-1}$ is the inverse of the 3 by 3 Gram matrix defined
by the momenta $p_1$, $p_2$, $p_3$. In terms of Mandelstam variables, 
these matrices are given by 
\bea
G = 
\left(\begin{array}{ccc} 0 & s & u \\ s & 0 & t \\ u & t & 0
\end{array}\right), &&\qquad 
H = \frac{1}{2}
\left(\begin{array}{ccc} 
{ 1/ s}+{ 1/ u} & { 1/ s} & { 1/ u} \\
{ 1/ s} & { 1/ s} +{ 1/ t}  &  { 1/ t} \\ 
{ 1/ u} & { 1/ t}       &  { 1/ t}+{ 1/ u}
\end{array}\right).
\eea 
${\cal P}$ is a projector onto the $(n-3)$ dimensional subspace
perpendicular to the 3 dimensional space spanned by the 
vectors $p_1$, $p_2$, $p_3$, $p_4$. ${\cal R}_{j,\nu}$ 
is the dual vector to $p_j^{\nu}$ relative to this 3 dimensional space.
One may easily check the following relations for the objects 
${\cal P}_{\mu\nu}$ and ${\cal R}_j^\mu$,
\bea
{\cal P}^{\mu\rho} {\cal P}_{\rho}^{\;\;\nu} &=& {\cal P}^{\mu\nu}, \label{P:1} \\
{\cal P}^{\mu}_{\;\;\nu} p_j^{\nu} &=& 0,\\
{\cal P}^{\mu}_{\;\;\mu} &=& tr({\cal P}) = tr({\cal P}\cdot {\cal P}) 
= n-3, \\
{\cal P}^{\mu}_{\;\;\nu} {\cal R}_j^{\nu} &=& 0,\\
{\cal R}_{j,\nu} p_k^{\nu} &=& \delta_{jk}, \\
{\cal R}_{j,\nu} {\cal R}_{l}^{\;\nu} &=& 2\, H_{jl}.\label{P:6}
\eea

To determine the tensor coefficients, $A_j$, 
 we define the following linear operators 
\bea\label{projectionsA}
\tilde A_1({\cal M}) &=& \frac{1}{(n-1)(n-3)} {\cal P}_{\mu_1\mu_2}{\cal P}_{\mu_3\mu_4} 
{\cal M}^{\mu_1 \mu_2\mu_3\mu_4},\nonumber \\
\tilde A_2({\cal M}) &=&  \frac{1}{(n-1)(n-3)} {\cal P}_{\mu_1\mu_3}{\cal P}_{\mu_2\mu_4} 
{\cal M}^{\mu_1 \mu_2\mu_3\mu_4},\nonumber \\
\tilde A_3({\cal M}) &=&  \frac{1}{(n-1)(n-3)} {\cal P}_{\mu_1\mu_4}{\cal P}_{\mu_2\mu_3} 
{\cal M}^{\mu_1 \mu_2\mu_3\mu_4}.
\eea
One finds by direct calculation that
\bea\label{atilde}
\left(\begin{array}{c}
  \tilde A_1({\cal M}) \\ \tilde A_2({\cal M}) \\ \tilde A_3({\cal M})  
\end{array}\right) &=&
\frac{tr({\cal P})}{(n-1)(n-3)} 
\left(\begin{array}{ccc}
  tr({\cal P}) & 1 & 1 \\  1 & tr({\cal P}) & 1 \\ 1 & 1 & tr({\cal P}) 
\end{array}\right)
\left(\begin{array}{c}
  A_1(s,t,u) \\  A_2(s,t,u) \\  A_3(s,t,u)  
\end{array}\right).
\eea
Staying in $n$-dimensions and inverting the system of equations yields,
\bea\label{ngoestofourproblem}
\left(\begin{array}{c}
  A_1(s,t,u) \\  A_2(s,t,u) \\  A_3(s,t,u)  
\end{array}\right)
 &=&
\frac{1}{(n-4)}
\left(\begin{array}{ccc}
  (n-2) & -1 & -1 \\  -1 & (n-2) & -1 \\ -1 & -1 & (n-2) 
\end{array}\right)
\left(\begin{array}{c}
  \tilde A_1({\cal M}) \\ \tilde A_2({\cal M}) \\ \tilde A_3({\cal M})  
\end{array}\right).
\eea
We see that the right hand side appears to introduce an 
additional factor of $1/(n-4)$. On the other hand, the sum
$$
A_1(s,t,u) + A_2(s,t,u) + A_3(s,t,u) = \tilde A_1({\cal M}) 
+ \tilde A_1({\cal M}) + \tilde A_1({\cal M})
$$
is free of spurious poles.
We will discuss how these apparent extra poles cancel
each other directly in the helicity
amplitudes below.    
 
It is straightforward to
define  projectors for the coefficients $B^{\alpha}_{lk}$,
\bea
\label{projectionsB}
\tilde B^{1}_{kl}({\cal M}) &=& \frac{1}{(n-3)} {\cal P}_{\mu_1\mu_2}
 {\cal R}_{k\mu_3}{\cal R}_{l\mu_4} {\cal M}^{\mu_1 \mu_2\mu_3\mu_4}
 \nonumber \\
 &&
 - \frac{2}{(n-4)} \, H_{kl}\, \left((n-2) \tilde A_1({\cal M})-\tilde A_2({\cal
 M})-\tilde A_3({\cal M}) \right),
\eea 
where the last term is proportional to $A_1$.
The other $B^{\alpha}$'s are found by permuting the 
Lorentz indices in (\ref{projectionsB}) 
and subtracting the respective $A$'s. 
Acting with Eq.~(\ref{projectionsB}) on the 
general tensor one finds the desired property,
$$
\tilde B^{\alpha}_{kl}({\cal M}) = B^{\alpha}_{kl}(s,t,u),
$$
by using relations (\ref{P:1}--\ref{P:6}).

In the same spirit one can construct projectors for the 
coefficients $C_{jklm}$, 
\bea\label{projectionsC}
\tilde C_{jklm}({\cal M}) &=& 
{\cal R}_{j\mu_1}{\cal R}_{k\mu_2}{\cal R}_{l\mu_3}{\cal R}_{m\mu_4} 
{\cal M}^{\mu_1\mu_2\mu_3\mu_4} \nonumber \\
&&- 2 \Bigl[  
  H_{jk} \tilde B^{1}_{lm}({\cal M}) 
+ H_{jl} \tilde B^{2}_{km}({\cal M}) 
+ H_{jm} \tilde B^{3}_{kl}({\cal M}) \nonumber\\
&&\hspace{0.35cm}+ H_{kl} \tilde B^{4}_{jm}({\cal M})
+ H_{km} \tilde B^{5}_{jl}({\cal M}) 
+ H_{lm} \tilde B^{6}_{jk}({\cal M}) \Bigr] \nonumber\\
&&- 4\, \Biggl[ 
 \frac{H_{jk}H_{lm}}{(n-4)} \, \left((n-2) \tilde A_1({\cal M})
                      -\tilde A_2({\cal M})-\tilde A_3({\cal M}) \right) \nonumber\\
&&\hspace{0.35cm}+\frac{H_{jl}H_{km}}{(n-4)} \, \left((n-2) \tilde A_2({\cal M})
                      -\tilde A_3({\cal M})-\tilde A_1({\cal M}) \right) \nonumber\\
&&\hspace{0.35cm}+\frac{H_{jm}H_{kl}}{(n-4)} \, \left((n-2) \tilde A_3({\cal M})
                      -\tilde A_1({\cal M})-\tilde A_2({\cal M}) \right)\Biggr], 
\eea
where the last three terms are proportional to $A_1$, $A_2$ and $A_3$
respectively and which again satisfies,
$$
\tilde C_{ijkl}({\cal M}) = C_{ijkl}(s,t,u).
$$

The linear operators acting on the amplitude defined in 
Eqs.~(\ref{projectionsA}),~(\ref{projectionsB}) and (\ref{projectionsC})  are
sufficient to determine any of the coefficients on the right hand side of the
general tensor decomposition of the amplitude in Eq.~(\ref{tensordec}).  The
linear nature ensures that   one can apply these projectors on a graph by graph
basis. 

All of the tensor coefficients  are linearly related to the basis set of
Eq.~(\ref{eq:basis}) by gauge invariance and permutation symmetry. An important
and non-trivial check of a full diagrammatic calculation is to evaluate each of
the tensor coefficients and to verify the relations amongst them imposed by the
Ward identities. 

It remains to be shown that the poles present in Eq.~(\ref{ngoestofourproblem})
cancel. This is seen by focussing on the dangerous $\tilde A_j$ terms only.
For $C_{2111}$ one finds that the $1/(n-4)$ pole drops out directly,
\bea
C_{2111}(s,t,u) &=& \tilde C_{2111}({\cal M}) \nonumber \\
&=& \dots + 4\,H_{11}\,H_{12} 
\Bigl[   A_1(s,t,u)+  A_2(s,t,u)+  A_3(s,t,u)\Bigr] .\nonumber 
\eea
For $B^1_{11}$ one has,
\begin{equation}
B^1_{11}(s,t,u) = \tilde B^1_{11}({\cal M}) = \dots +  \frac{t}{su}\,
A_1(s,t,u).  
\end{equation}
Using the Bose symmetry relations,
$$
A_1(s,t,u)=A_1(s,u,t)\, , \, A_2(s,t,u)=A_1(u,t,s)\, , \, A_3(s,t,u)=A_1(t,s,u) \, ,
$$
a short calculation shows that in   
Eqs.~(\ref{ampPPPP}),~(\ref{ampPPPM}) and (\ref{ampPPMM}) only
the sum of the $A_j$'s is present so that the helicity amplitudes
are free of spurious poles.
This ensures that
the projector method can be applied on a graph by graph basis without
the need to expand one order higher in $\epsilon$.

\section{Light-by-light scattering in SUSY QED}
\label{sec:SUSYamps}

\subsection{The SUSY Ward Identity}
\label{subsec:SWI}

Supersymmetric amplitudes are expected to satisfy the Supersymmetric Ward
identity~\cite{Grisaru:1977vm,Grisaru:1977px,Parke:1985pn},
\begin{equation}
0 = \langle 0 | ~[ Q,a_1^\dagger\ldots a_n^\dagger ] ~ | 0\rangle,
\label{eq:swi}
\end{equation}
where $Q$ is the Supersymmetry generator that satisfies $Q | 0\rangle = 0$ and
$a_i^\dagger$ are the creation operators for particles in the initial state.
Application of Eq.~(\ref{eq:swi}) to states with three 
photons and a photino yields the following identities,
\begin{eqnarray}
\label{eq:swipppp}
{\cal M}_{++++} &\equiv& 0,\\
\label{eq:swipppm}
{\cal M}_{+++-} &\equiv& 0.
\end{eqnarray}
At tree level these identities are trivially satisfied due to the
abelian nature of the photon.   Beyond tree-level, couplings with
the matter multiplet give rise to non-trivial interactions, that we
investigate in the subsequent subsections.  
As mentioned in the introduction, we work in conventional dimensional
regularisation and treat all momenta and Lorentz indices as
$n$-dimensional.  This breaks supersymmetry because the fermionic and
bosonic degrees of freedom for the gauge multiplet are no longer
equivalent.   We therefore expect that there will be non-trivial
corrections to the SUSY Ward identity that vanish in the 4-dimensional limit. 

For convenience we decompose the helicity amplitudes perturbatively as
\begin{equation}
{\cal M}_{\lambda_1\lambda_2\lambda_3\lambda_4}
= 
\alpha^2 \left( \M{1}{\lambda_1\lambda_2\lambda_3\lambda_4} 
+ \frac{\alpha}{\pi} \M{2}{\lambda_1\lambda_2\lambda_3\lambda_4} 
+ {\cal O}(\alpha^2)\right).
\end{equation}

\subsection{One-loop SUSY QED helicity amplitudes}
\label{subsec:oneloop}

Because of the abelian nature of QED, there is no quartic photon coupling at tree level.
At one-loop a four-point contribution is generated by the interaction of the photon with the
matter multiplet.
The one-loop amplitude can be decomposed according to gauge invariant subsets
of graphs.   There are two such groups at one-loop
so that
\begin{equation}
\M{1}{ }
=\MS{1}{ }
+\MF{1}{ },
\end{equation}
where the dependence on the helicities has been suppressed.
These two contributions, $\MS{1}{}$ and $\MF{1}{}$,  denote the 
scalar electron  and electron loops respectively.
In the Standard Model, only the electron loops, $\MF{1}{}$, contribute.
We find
\begin{eqnarray}
\MS{1}{++++} &=& \phantom{-}8 + {\cal O}(\epsilon),\nonumber \\
\MF{1}{++++} &=& -8+ {\cal O}(\epsilon),\nonumber \\
\MS{1}{+++-} &=& -8+ {\cal O}(\epsilon),\nonumber \\
\MF{1}{+++-} &=& \phantom{-}8+ {\cal O}(\epsilon),\nonumber \\
\MS{1}{++--} &=& 4 - 2 \left((X-Y)^2+\pi^2\right) 
+ 4 \left( (X-Y)^2 + \pi^2 -2 (X-Y)\right) \tttu \nonumber \\
&&+ \Biggl \{ u \leftrightarrow t \Biggr \}+ {\cal O}(\epsilon), \nonumber \\  
\MF{1}{++--} &=& -4 
- 4 \left( (X-Y)^2 + \pi^2 -2 (X-Y)\right) \tttu \nonumber \\
&&+ \Biggl \{ u \leftrightarrow t \Biggr \} + {\cal O}(\epsilon), 
\label{eq:oneconts}
\end{eqnarray}
where 
\begin{equation}
X = \log\left(\frac{-t}{s}\right),
\qquad Y = \log\left(\frac{-u}{s}\right).
\end{equation}

Combining all graphs together, the one-loop light-by-light scattering amplitudes
in SUSY QED are rather compact and are given by,
\begin{eqnarray}
\label{eq:oneresult}
\M{1}{++++}&\equiv&0 \nonumber \\
\M{1}{+++-}&\equiv&0\nonumber \\
\M{1}{++--}&\equiv& 4s (n-4) \left( {\rm Box}^6(s,t)+{\rm Box}^6(s,u)\right)
-4s(n-2){\rm Box}^6(t,u)\nonumber \\
&=&-4 \left( (X -Y)^2 + \pi^2\right) + {\cal O}(\epsilon),
\end{eqnarray}
where ${\rm Box}^6(s,t)$ is the (infrared and ultraviolet finite)
one-loop box graph in $6-2\epsilon$ dimensions.  At one-loop there is no contribution
from the gauge multiplet and this is therefore the result for both $N=1$ and $N=2$ SUSY QED.

The fact that $\M{1}{++++}$ and $\M{1}{+++-}$ both vanish is directly
attributable to the SUSY Ward identity.   The zeroes for $\M{1}{++++}$ and
$\M{1}{+++-}$ occur at the level of the master integrals, i.e. to all
orders in $n-4$.   This is perhaps not surprising since although 
dimensional regularisation breaks the supersymmetry for the gauge
multiplet, the photon and photino are not present as internal particles in
any of the one-loop graphs.  At this order, we are not sensitive to the SUSY
breaking engendered by dimensional regularisation.

At one-loop we expect that amplitudes contain  terms of weight 0,
1 and 2 (counting logarithms and $\pi$ as weight 1, squares of logarithms as
weight 2 and constants (excluding $\pi$) as weight 0).   By inspection of 
Eq.~(\ref{eq:oneconts}), we see that all such terms are present in the fermion
and scalar contributions.  However, the supersymmetric cancellation is such that
only terms of weight 2 remain  in Eq.~(\ref{eq:oneresult}), all weight 0 and 
weight 1 contributions are eliminated.
We also note that terms proportional to 
ratios of the kinematic scales $t^2/s^2$ in the individual contributions
(\ref{eq:oneconts}) completely cancel
in the supersymmetric result of Eq.~(\ref{eq:oneresult}).

\subsection{Two-loop SUSY QED helicity amplitudes}
\label{subsec:twoloop}

\FIGURE[t]{
\label{fig:photinoloop}
\begin{tabular}{cc}
\scalebox{.3}{\includegraphics{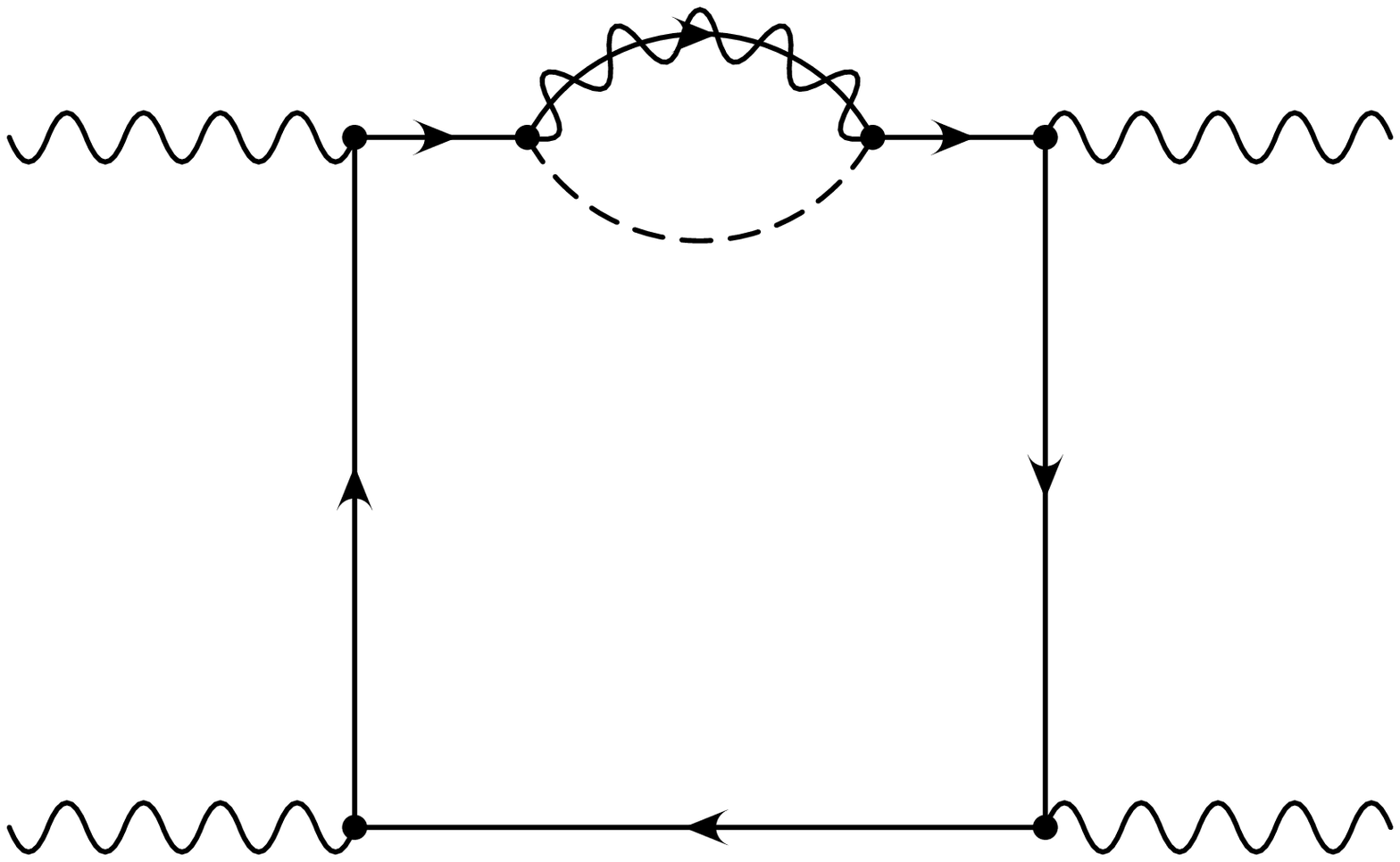}}
\scalebox{.3}{\includegraphics{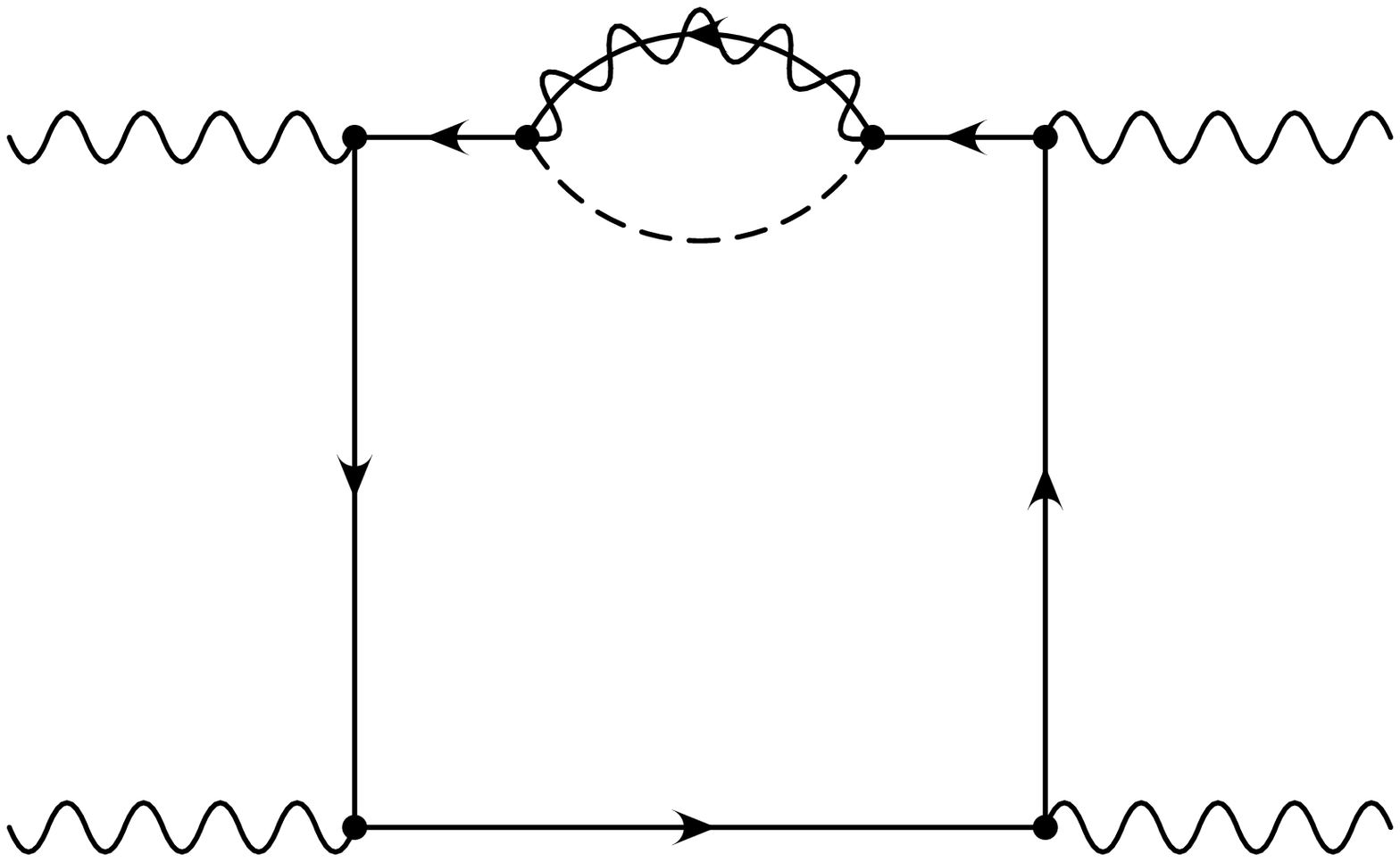}}
\end{tabular}
\caption{Typical diagrams involving the photino that are related by charge
conjugation. The dashed line represents the scalar electron, while the 
overlaid solid and wavy lines
represent the photino.  Similar graphs can be drawn where the 
dashed line represents the scalar photon and the overlaid solid and wavy lines
represent the electron.}
} 

Unlike the one-loop case, the chiral photino-electron-selectron coupling
is present and it is necessary to address the issue of how to treat
$\gamma_5$. Because of the parity invariance of the process, the scattering
amplitude contains no antisymmetric part, indicating that the $\gamma_5$
contributions drop out.  We can see this by considering a typical pair of
diagrams with a photino exchange that are related by charge conjugation as
shown in Fig.~\ref{fig:photinoloop}.  Up to overall factors, the
contribution from the first diagram is given by
\begin{equation}
I_1 \sim \int \frac{d^nk}{(2\pi)^n} \frac{d^n\ell}{(2\pi)^n}
\frac{Tr(P_R \sss{\ell} P_L
\sss{k}\varepsilon_1\sss{k_{1}}\varepsilon_2\sss{k_{12}}\varepsilon_3\sss{k_{123}}\varepsilon_4\sss{k}) }{(k^2)^2(k_1)^2(k_{12})^2(k_{123})^2 \ell^2 (k-\ell)^2}
\end{equation}
where $k_{i\ldots j} = k+ p_i +\ldots+p_j$ and 
\begin{equation}
P_L = \PL, \qquad P_R = \PR.
\end{equation}
Similarly the second graph is  
\begin{equation}
I_2 \sim \int \frac{d^nk}{(2\pi)^n} \frac{d^n\ell}{(2\pi)^n}
\frac{Tr(P_R \sss{\ell} P_L
\sss{k}\varepsilon_4\sss{k_{4}}\varepsilon_3\sss{k_{34}}\varepsilon_2\sss{k_{234}}
\varepsilon_1\sss{k}) }{(k^2)^2(k_4)^2(k_{34})^2(k_{234})^2 \ell^2 (k-\ell)^2}.
\end{equation}
Relabelling $k \to -k$, $\ell \to -\ell$ and using charge conjugation to reverse
the trace, we find that,
\begin{equation}
I_2 \sim \int \frac{d^nk}{(2\pi)^n} \frac{d^n\ell}{(2\pi)^n}
\frac{Tr(P_L \sss{\ell} P_R
\sss{k}\varepsilon_1\sss{k_{1}}\varepsilon_2\sss{k_{12}}\varepsilon_3\sss{k_{123}}\varepsilon_4\sss{k}) }{(k^2)^2(k_1)^2(k_{12})^2(k_{123})^2 \ell^2 (k-\ell)^2}
\end{equation}
so that, up to common factors 
\begin{equation}
I_1+I_2  \sim  \left[ P_R \sss{\ell} P_L + P_L \sss{\ell} P_R \right]\ldots  =
\frac{1}{2}\left[ \sss{\ell} -\gamma_5 \sss{\ell}\gamma_5 \right] \ldots. 
\end{equation}
As expected, traces with single $\gamma_5$ factors drop out entirely.   Since
the amplitudes are finite, we therefore use an anti-commuting $\gamma_5$ such
that,
\begin{equation}
\frac{1}{2}\left[ \sss{\ell} -\gamma_5 \sss{\ell}\gamma_5 \right] = \sss{\ell}.
\end{equation}
Similar arguments apply to the chiral couplings of the scalar photon with the electron.

\subsubsection{$N=1$ SUSY QED}

As in the one-loop case, it is convenient to 
break the amplitude up according to
the different gauge-invariant pieces so that
\begin{equation}
\label{eq:N1sum}
\M{2}{}
=\MS{2}{}
+\MF{2}{}
+\MP{2}{}
+\MV{2}{},
\end{equation}
where the dependence on the helicities has been suppressed.
At two-loops there are contributions
from photino exchange, $\MP{2}{}$, and the four scalar vertex, $\MV{2}{}$,
as well as graphs where the electron or
scalar electron couple directly to the photons, $\MF{2}{}$ and $\MS{2}{}$.  
 Altogether
there are 1902 Feynman graphs which we generate using QGRAF\cite{Nogueira:1991ex}.  
Figures~\ref{fig:A2},~\ref{fig:B2},~\ref{fig:C2},~\ref{fig:D2} and ~\ref{fig:E2} show the 
types of Feynman graphs relevant for the various contributions.
Tadpole graphs and self-energies of external legs are not shown since they
vanish in dimensional regularisation. 
We note that all of the possible diagrams are planar.
\FIGURE[t]{
\label{fig:A2}
\begin{tabular}{ccc}
\scalebox{.2}{\includegraphics{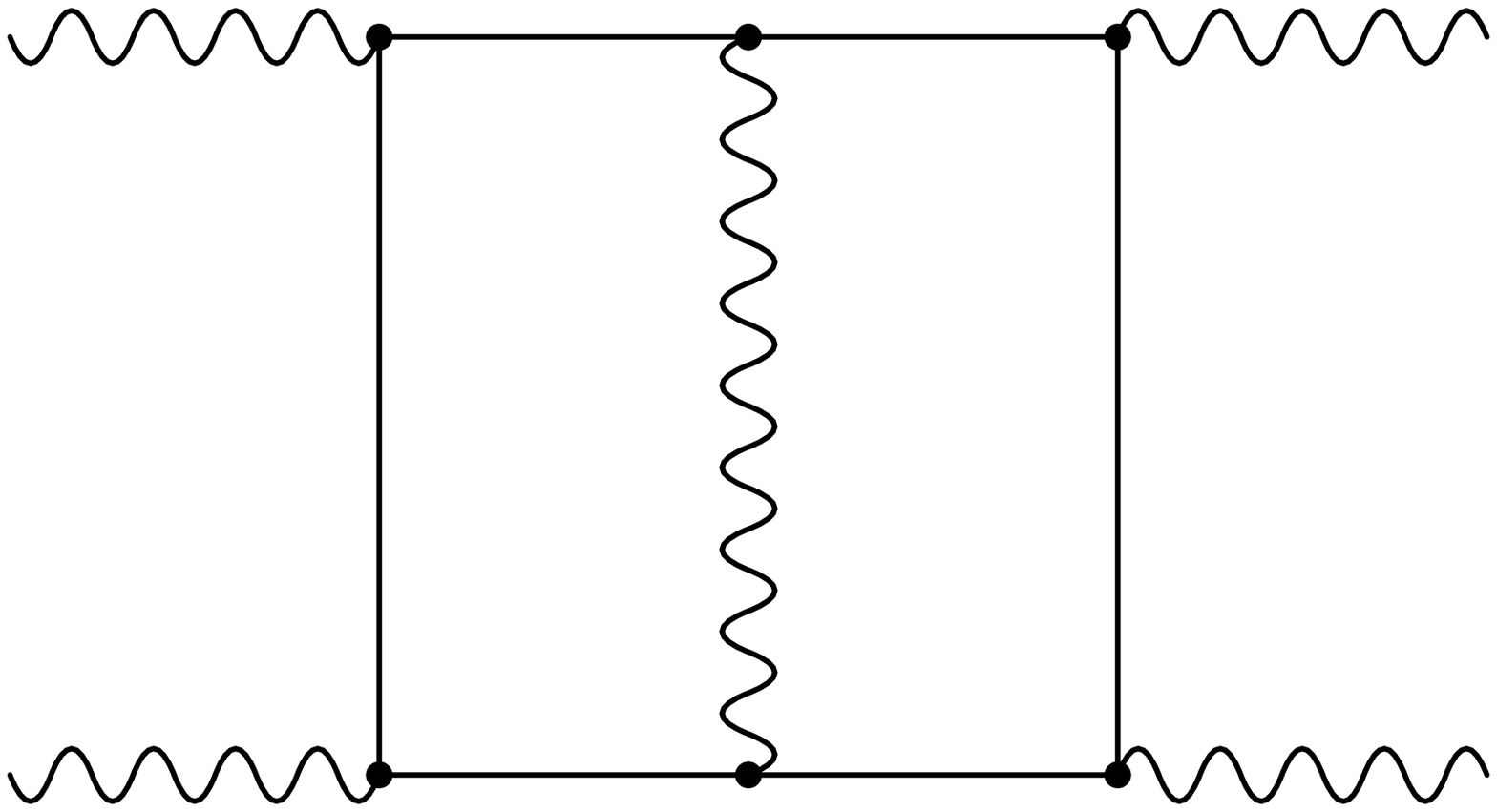}}&
\scalebox{.2}{\includegraphics{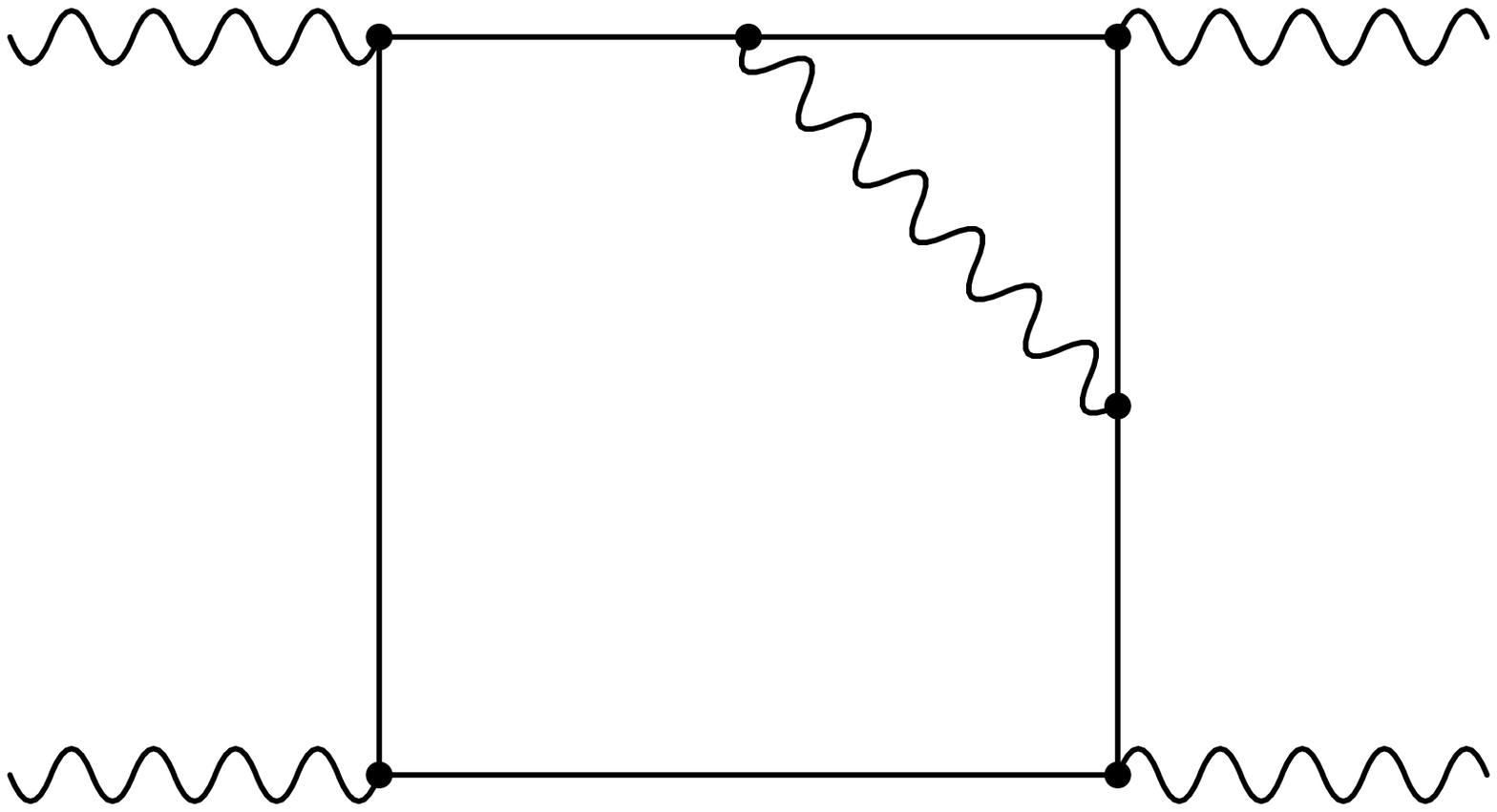}}&
\scalebox{.2}{\includegraphics{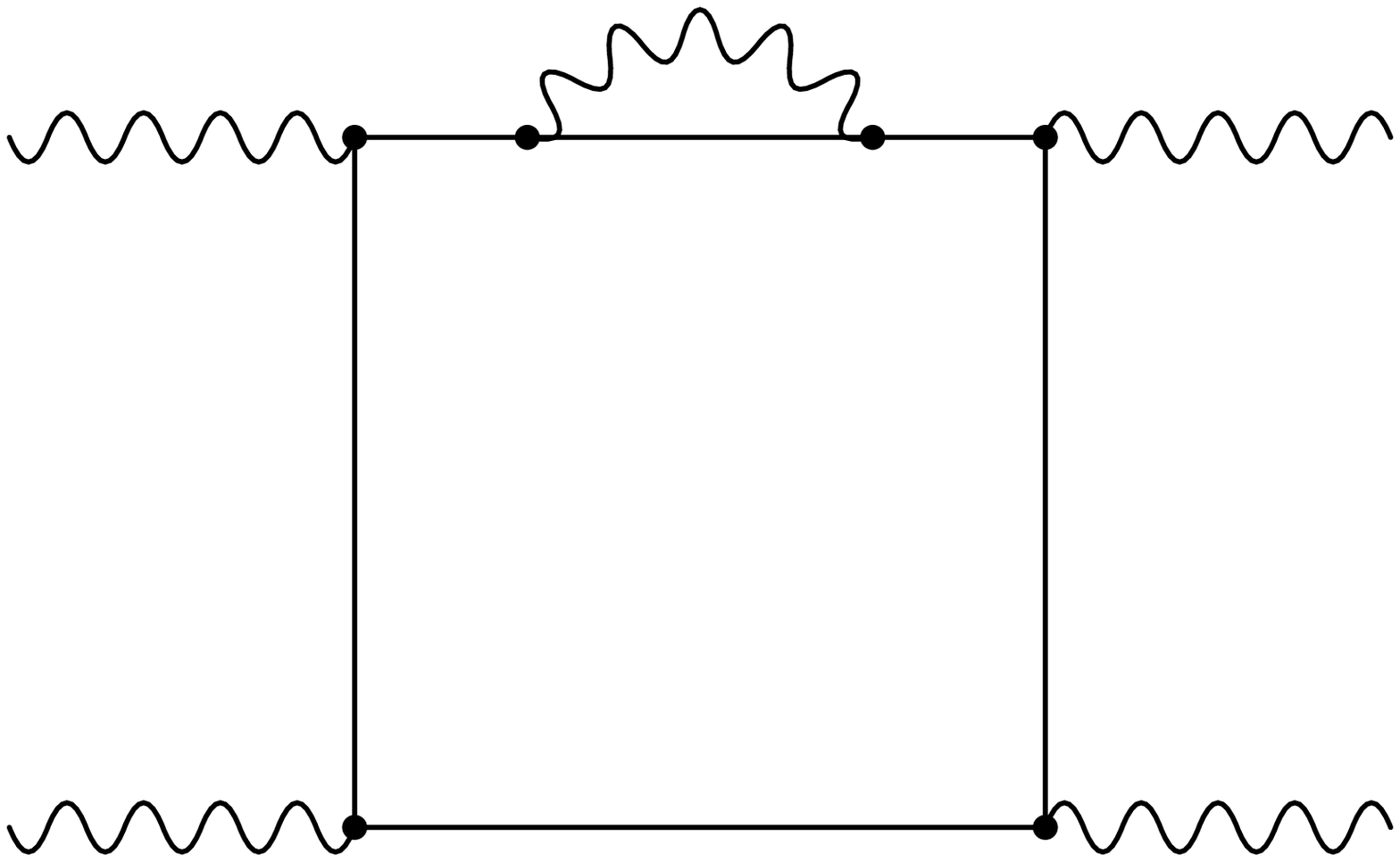}}\\
A&B&C
\end{tabular}
\caption{Graphs relevant for the electron, photino, scalar electron and scalar photon
contributions
$\MF{2}{}$, $\MP{2}{}$, $\MS{2}{}$ and $\MX{2}{}$. 
The solid lines represent particles from the matter multiplets, i.e. the electron
or scalar electron while the internal wavy lines
denote particles from the gauge multiplets, the photon, the photino or the scalar photon.}
}
\FIGURE[h]{
\label{fig:B2}
\begin{tabular}{ccc}
\scalebox{.2}{\includegraphics{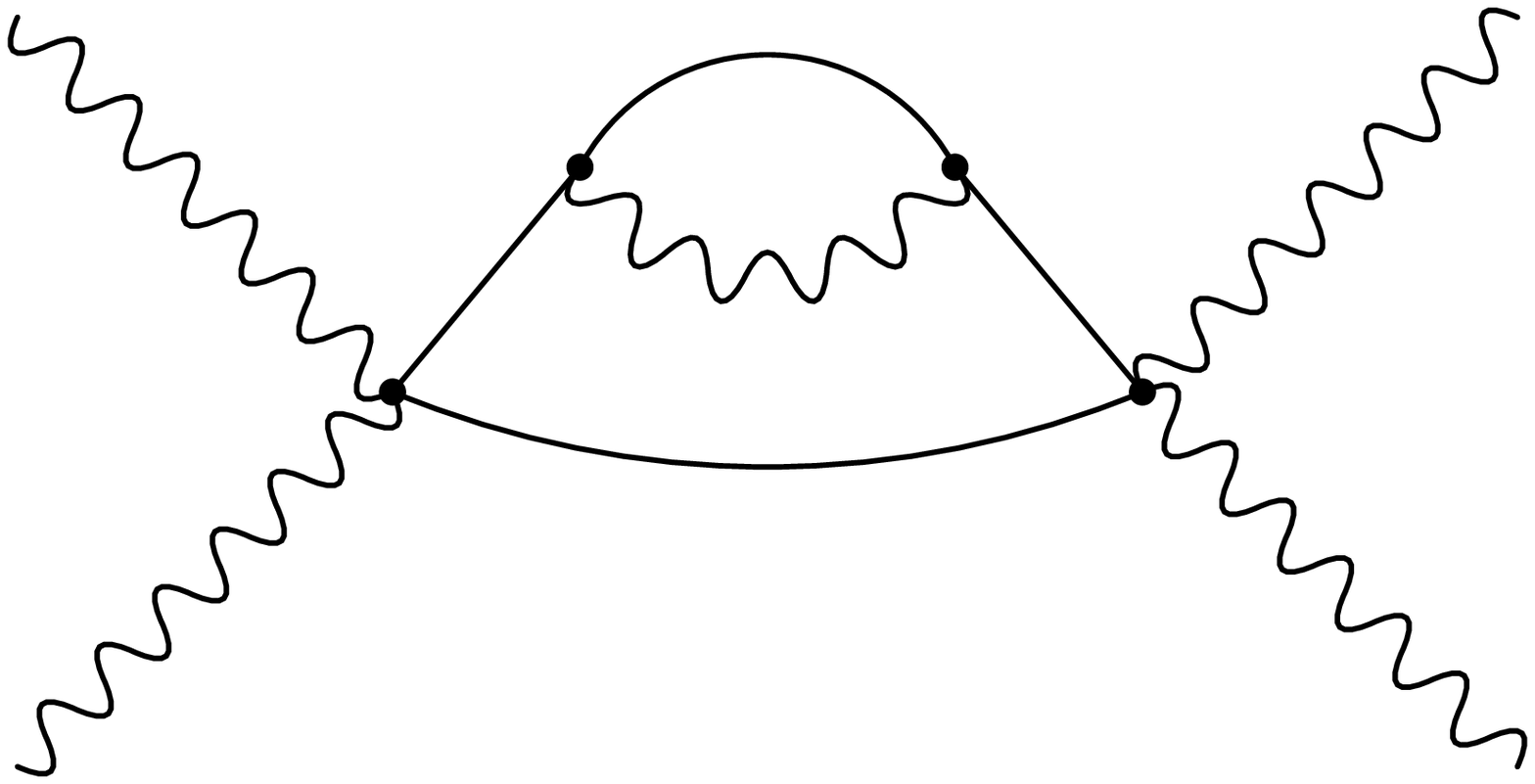}}&
\scalebox{.2}{\includegraphics{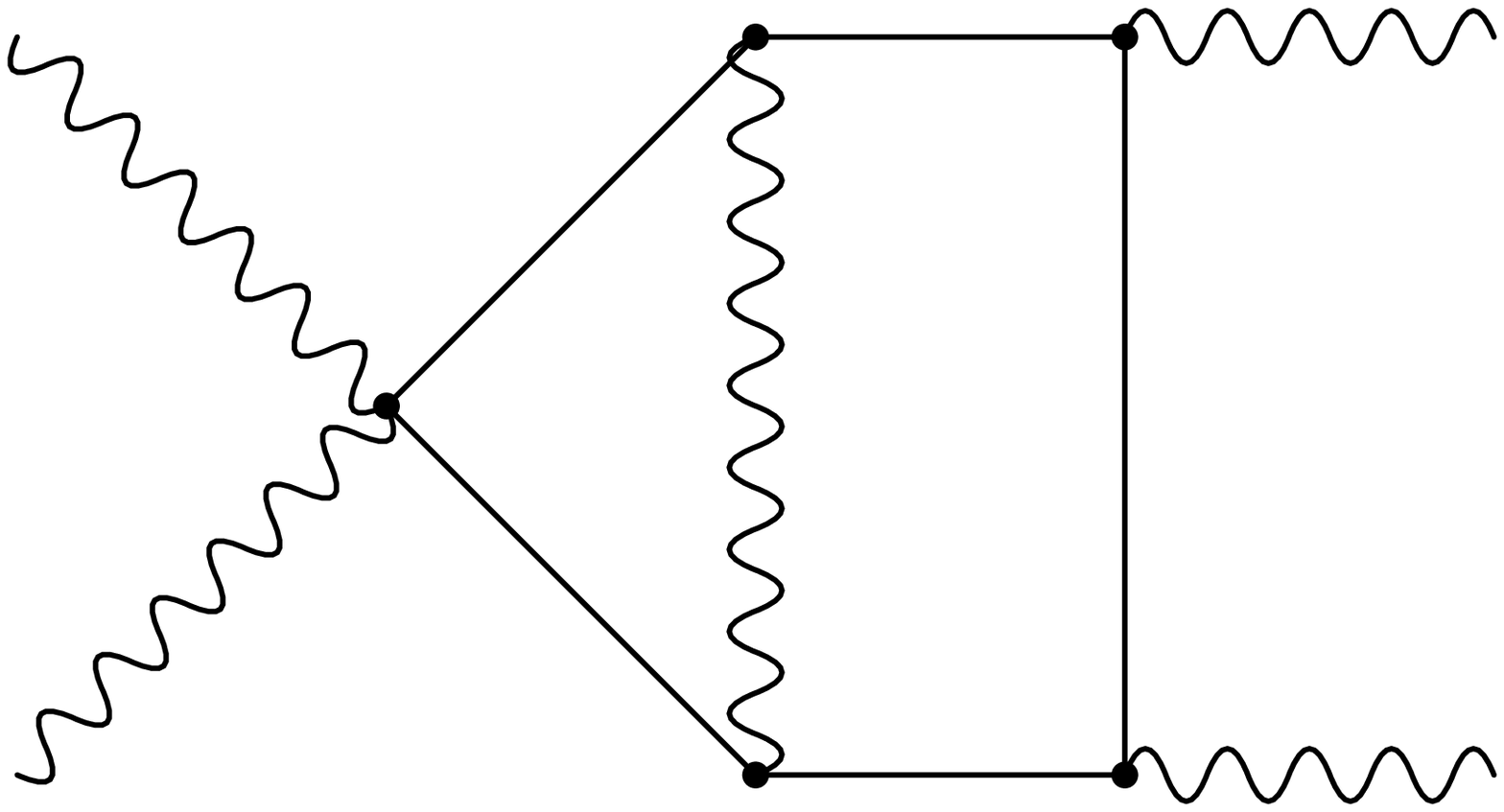}}&
\scalebox{.2}{\includegraphics{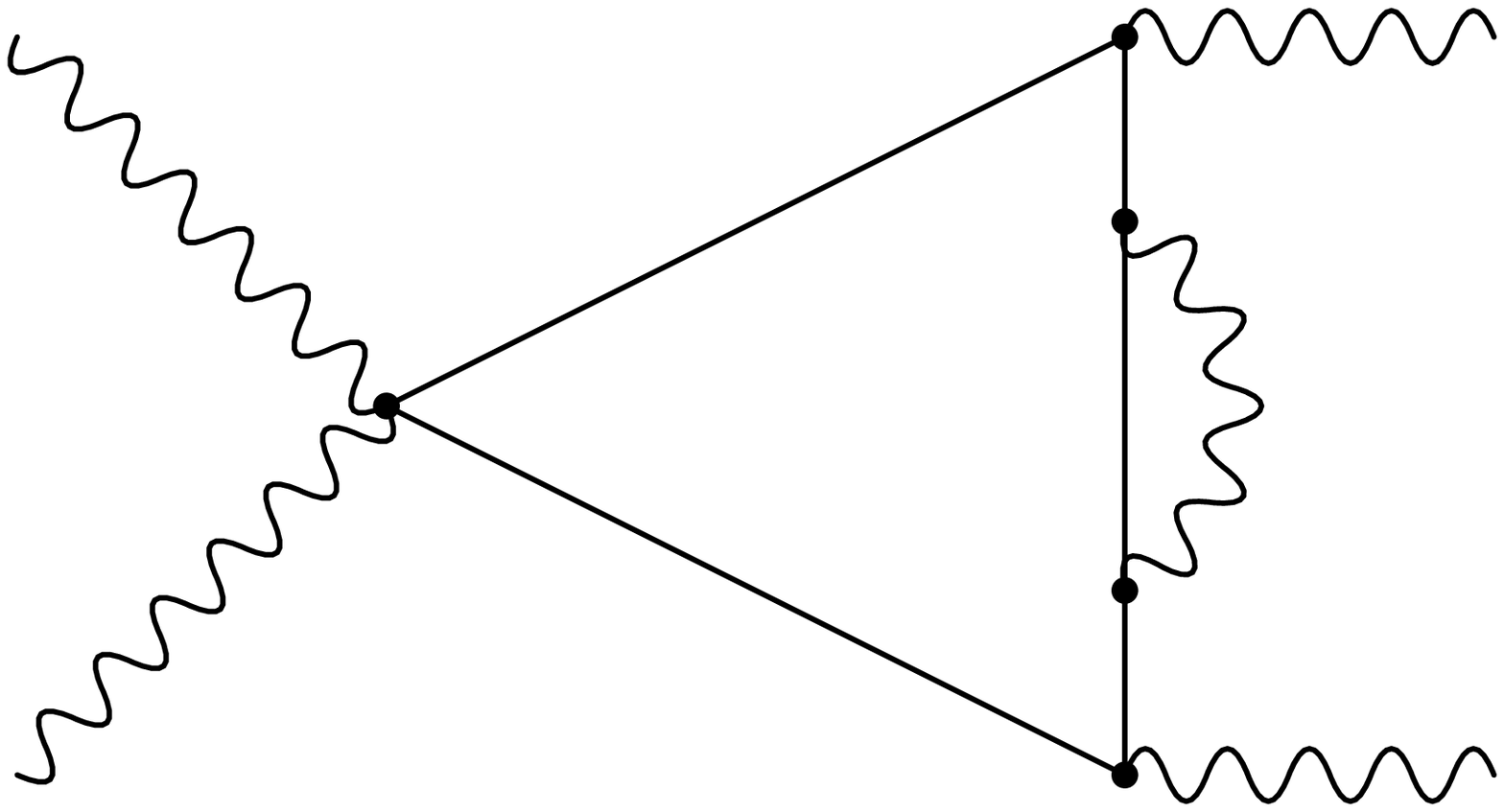}}\\
D&E&F\\
\scalebox{.2}{\includegraphics{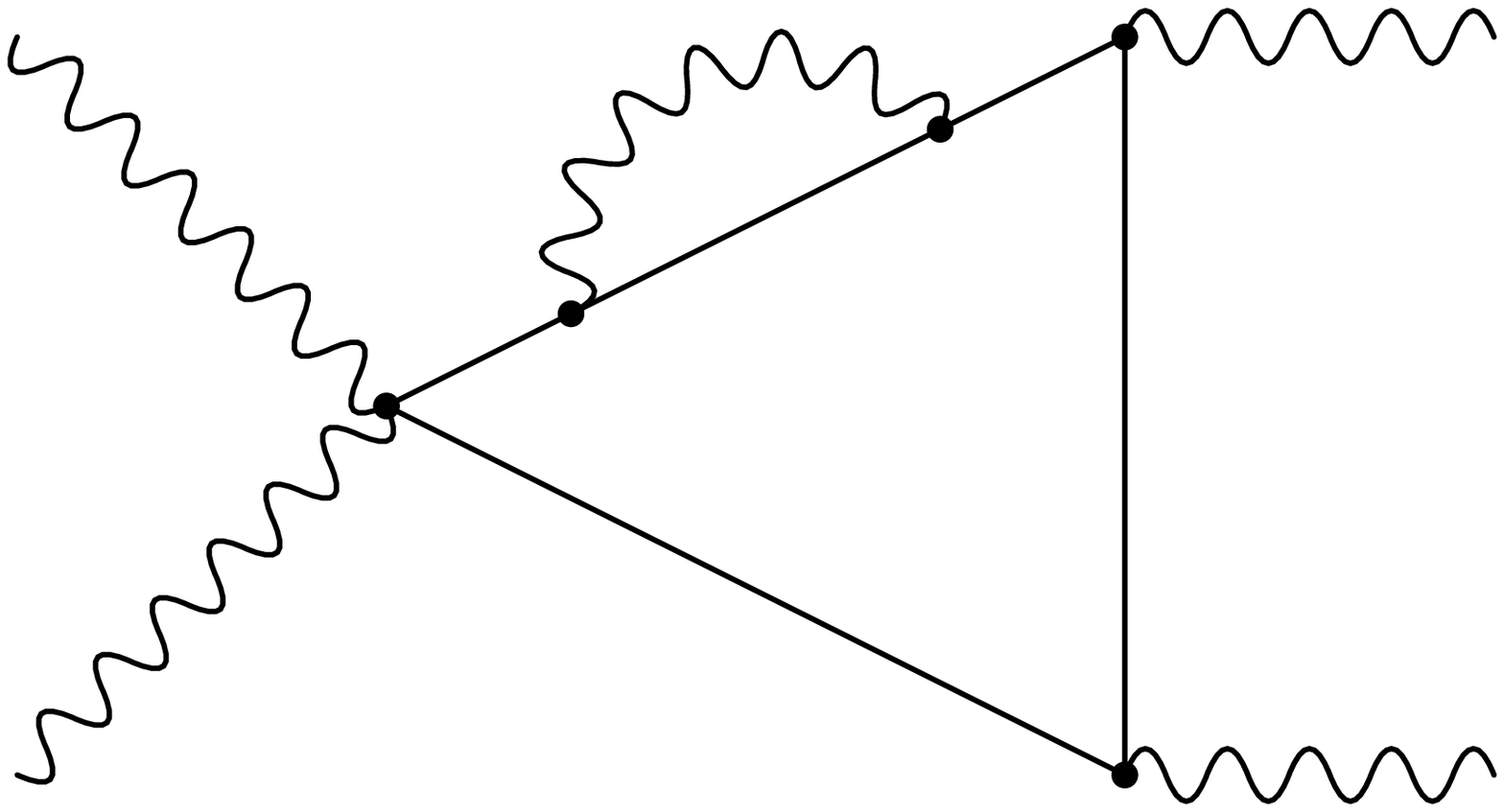}}&
\scalebox{.2}{\includegraphics{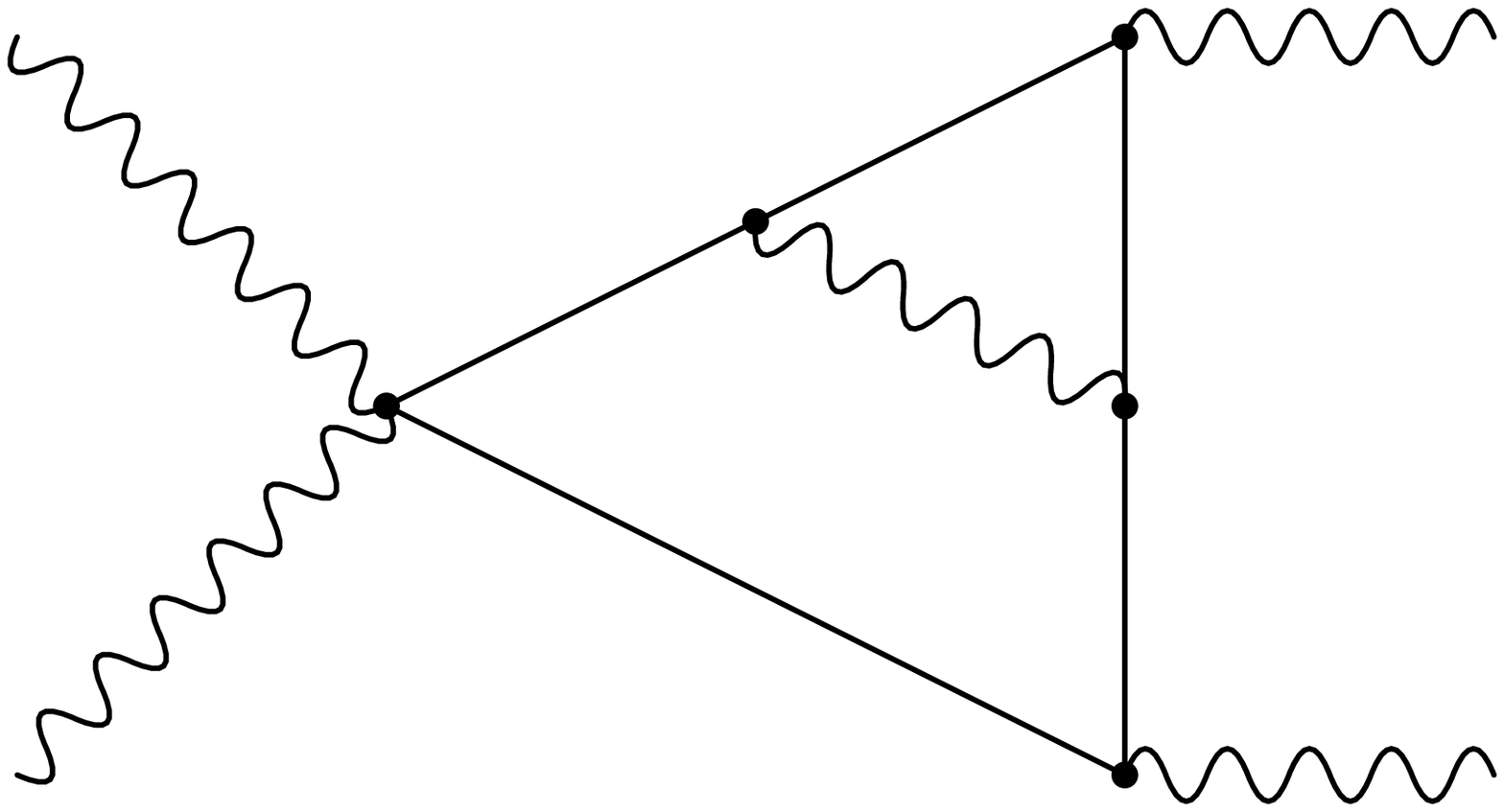}}\\
G&H
\end{tabular}
\caption{Graphs relevant for the photino and scalar electron contributions
$\MP{2}{}$ and $\MS{2}{}$.
The solid lines represent the scalar electron while the wavy lines denote either the photon or photino.}
}
\FIGURE[h]{
\label{fig:C2}
\begin{tabular}{ccc}
\scalebox{.2}{\includegraphics{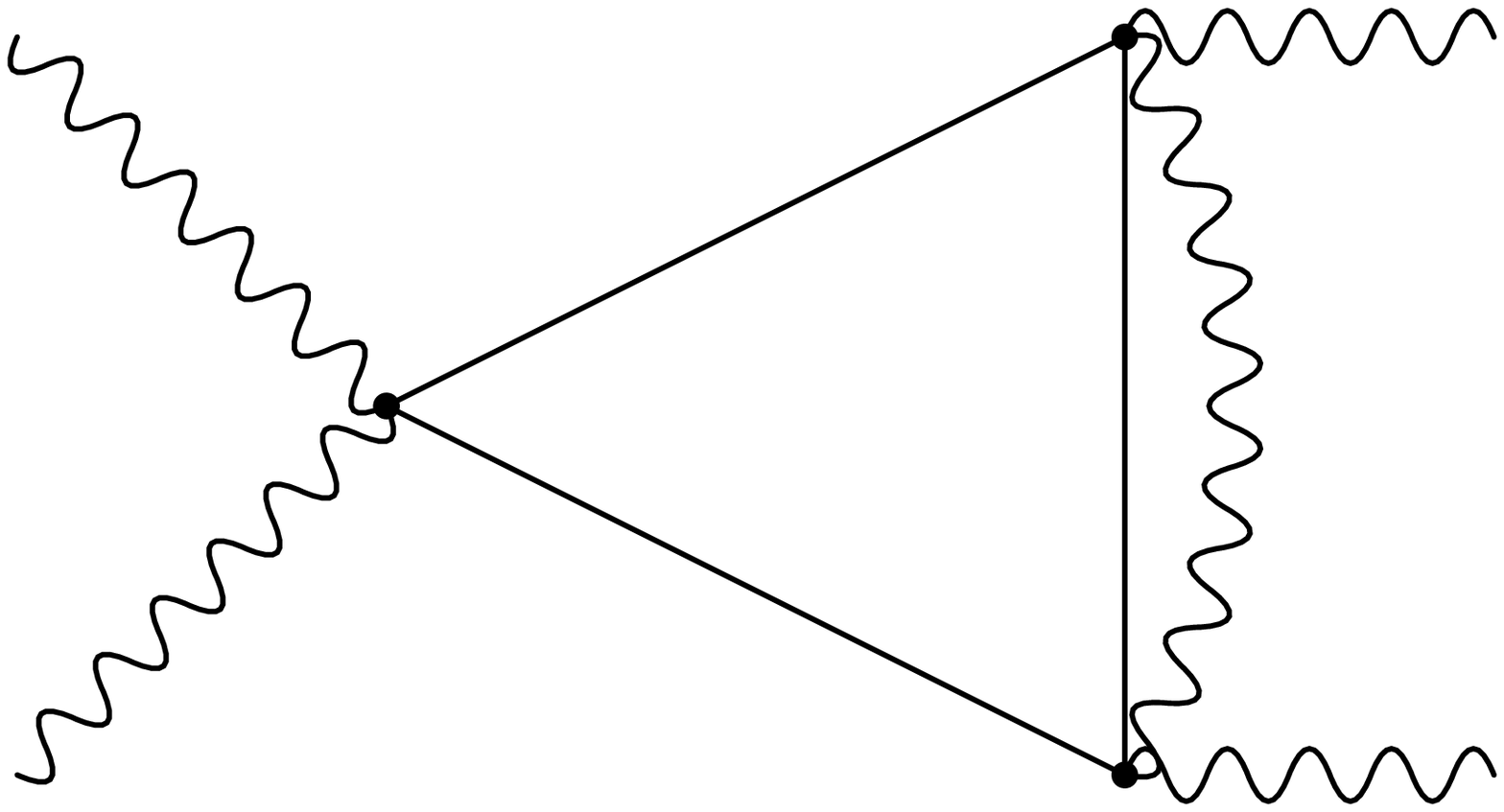}}&
\scalebox{.2}{\includegraphics{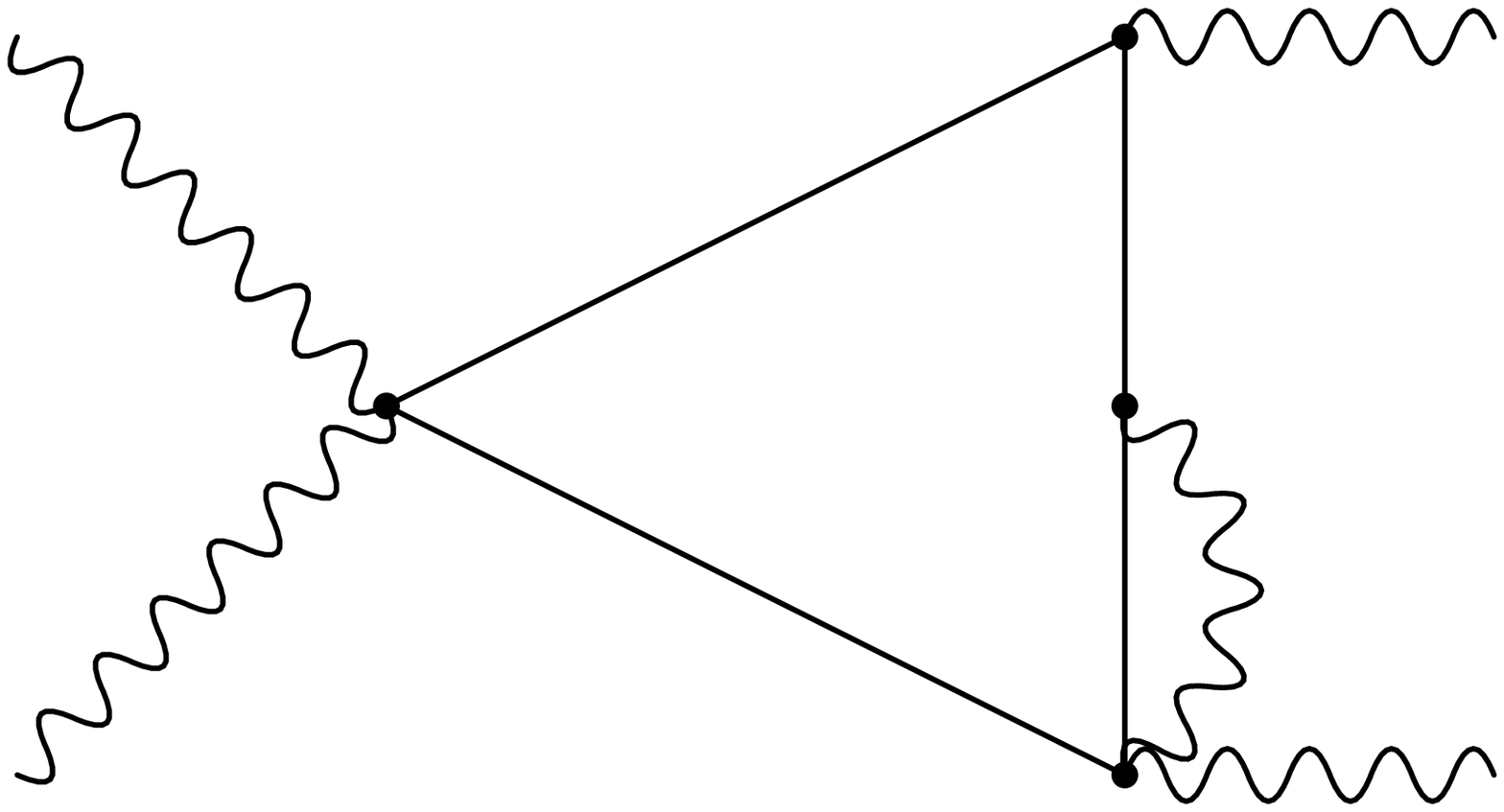}}&
\scalebox{.2}{\includegraphics{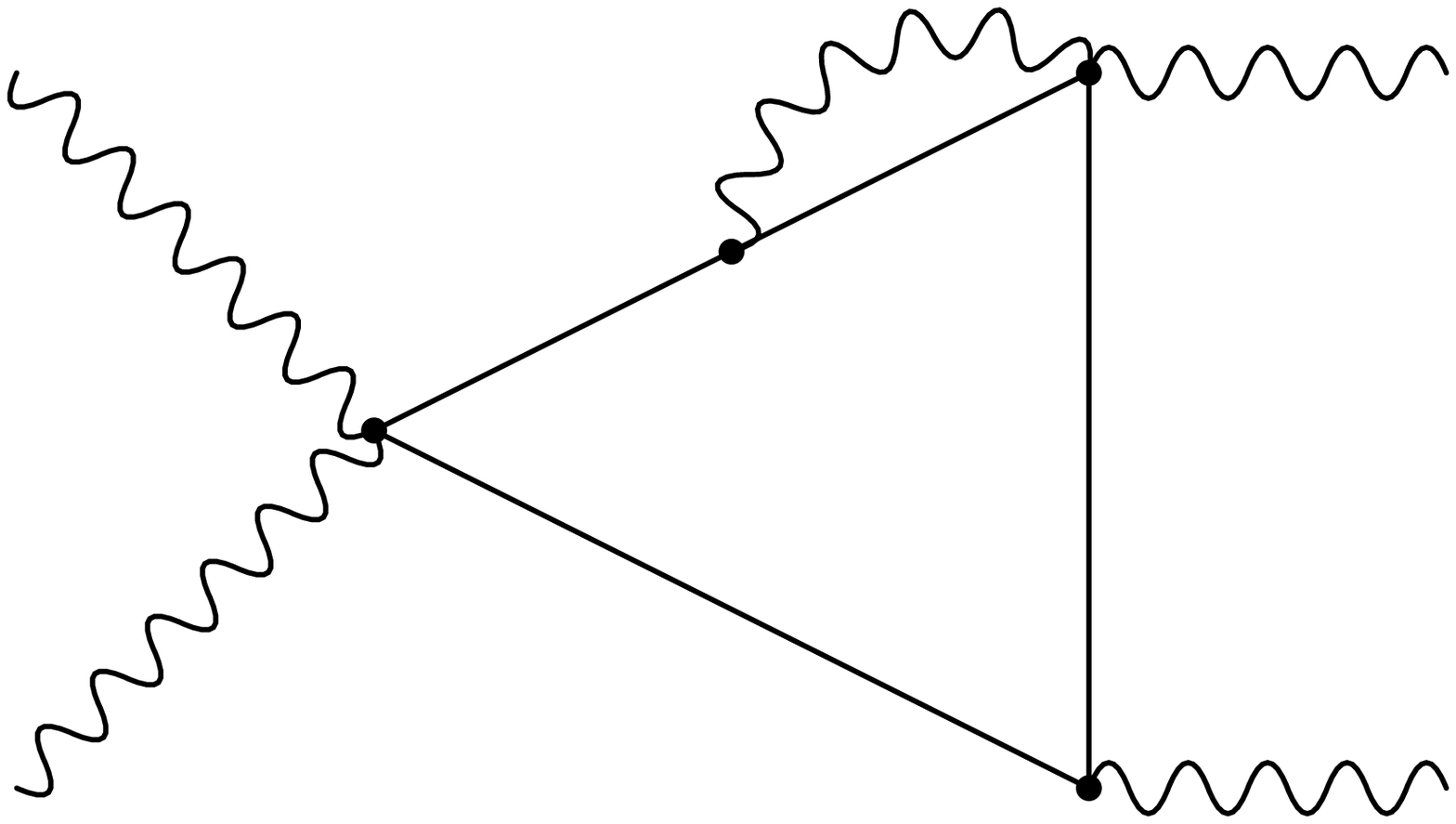}}\\
I&J&K\\
\scalebox{.2}{\includegraphics{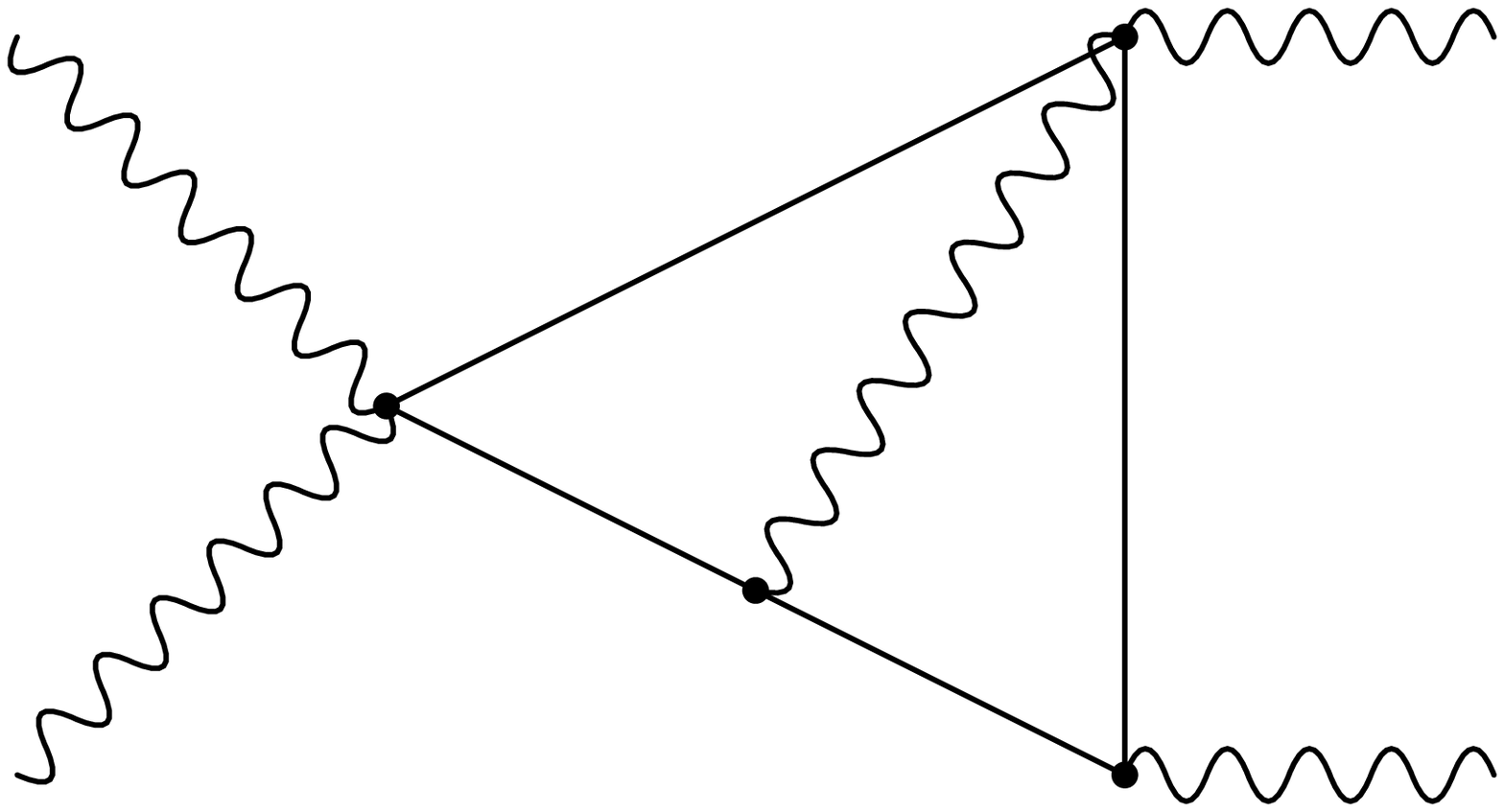}}&
\scalebox{.2}{\includegraphics{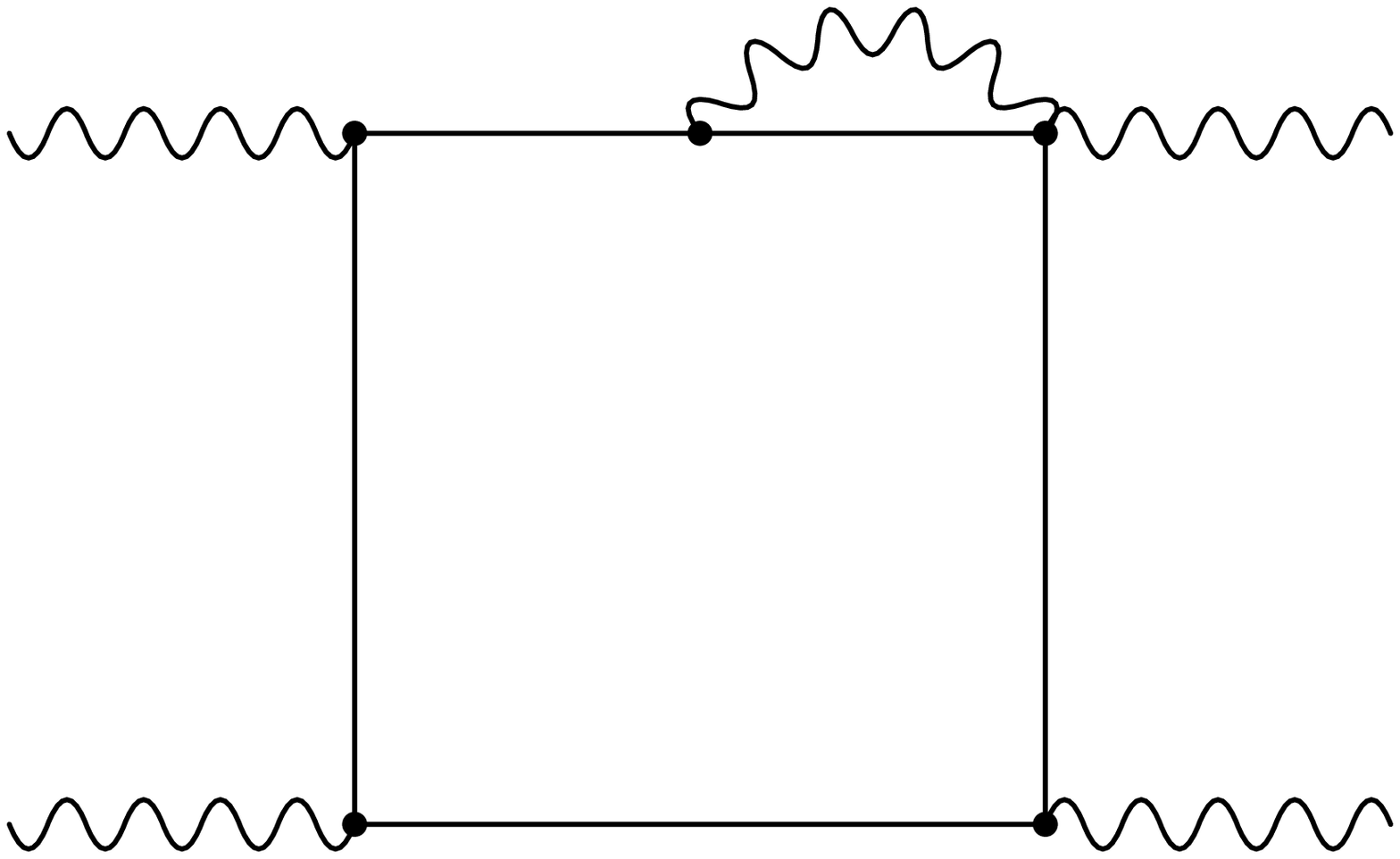}}&
\scalebox{.2}{\includegraphics{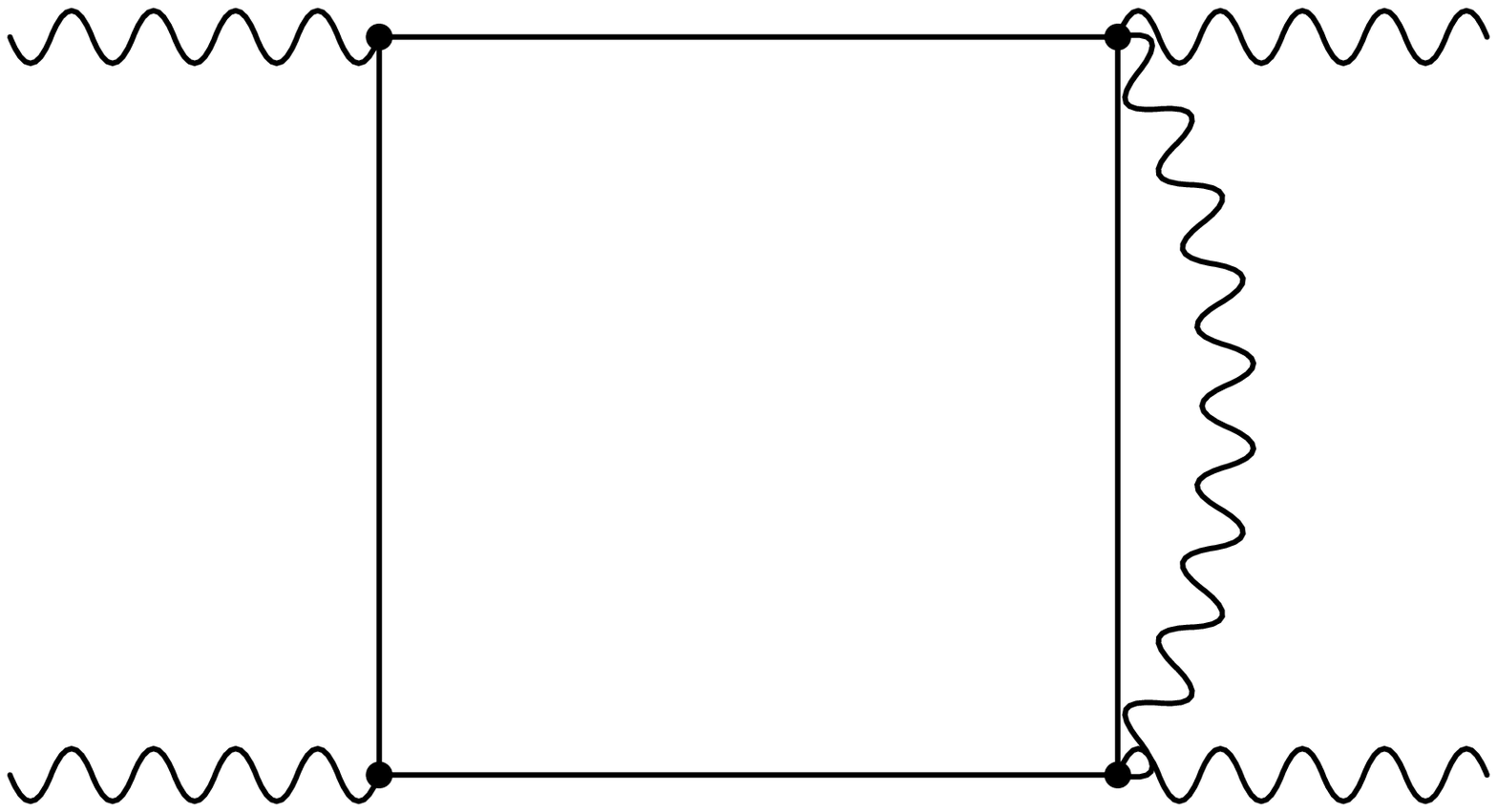}}\\
L&M&N\\
\scalebox{.2}{\includegraphics{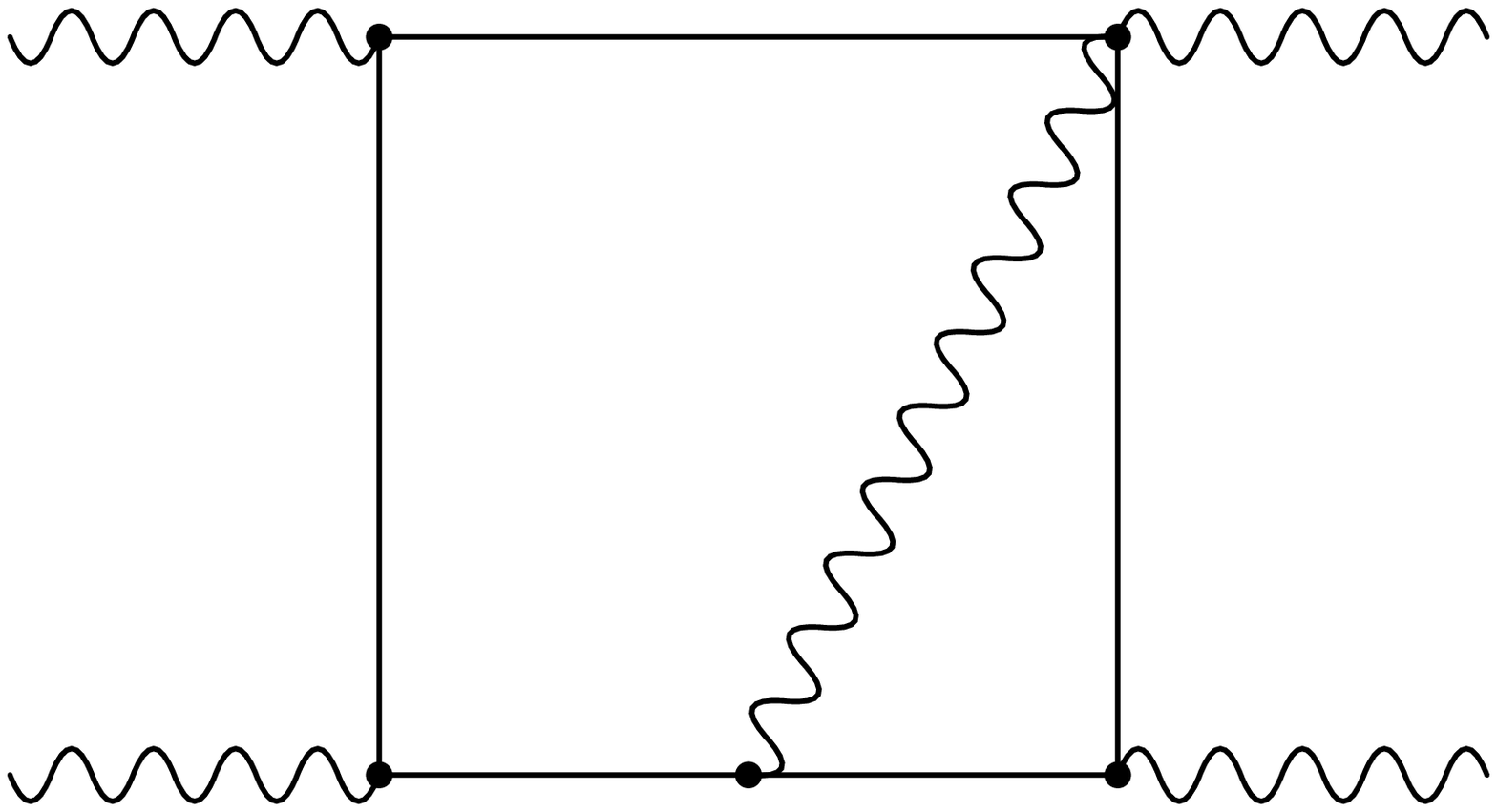}}&
\scalebox{.2}{\includegraphics{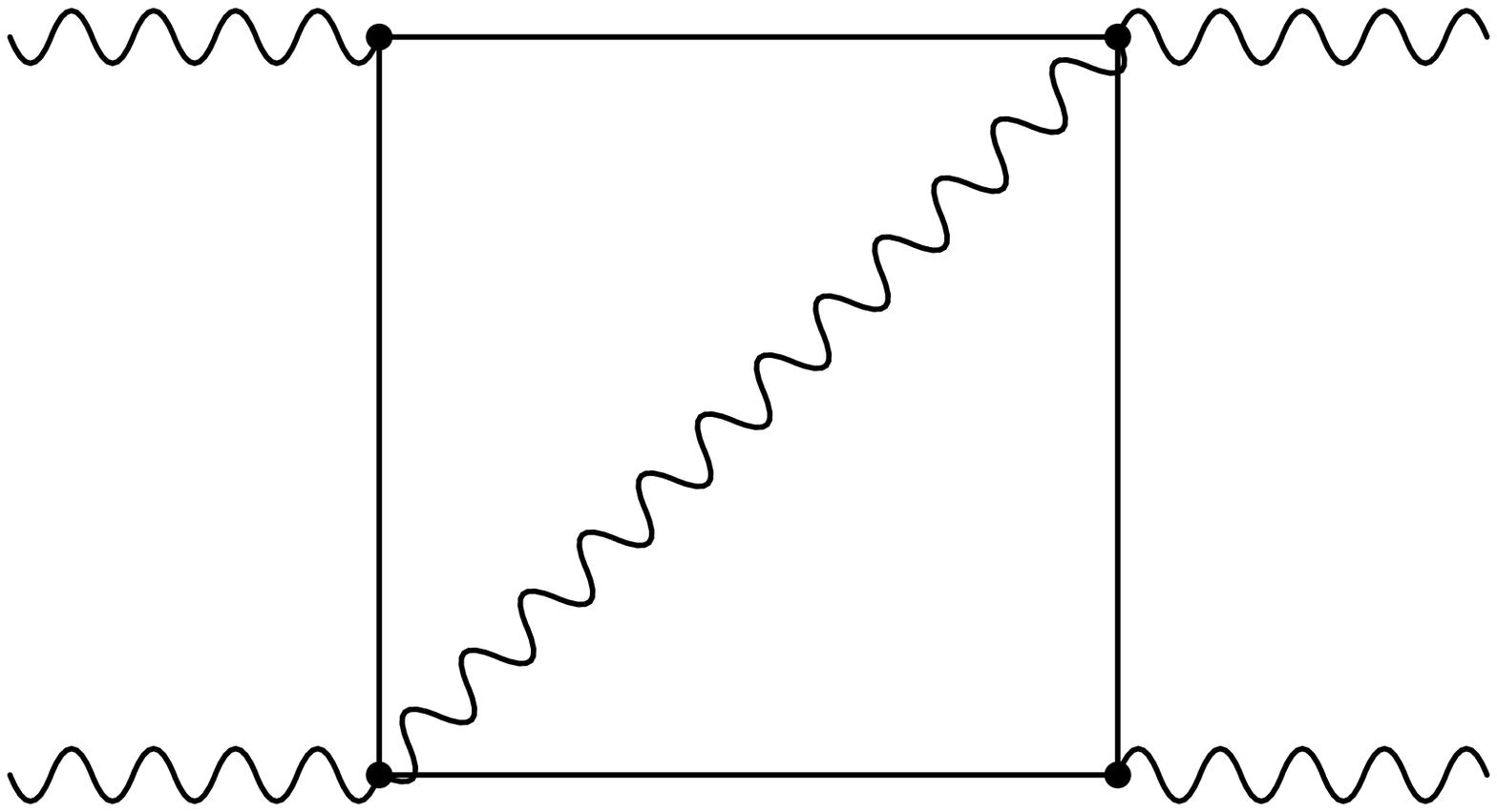}}&
\scalebox{.2}{\includegraphics{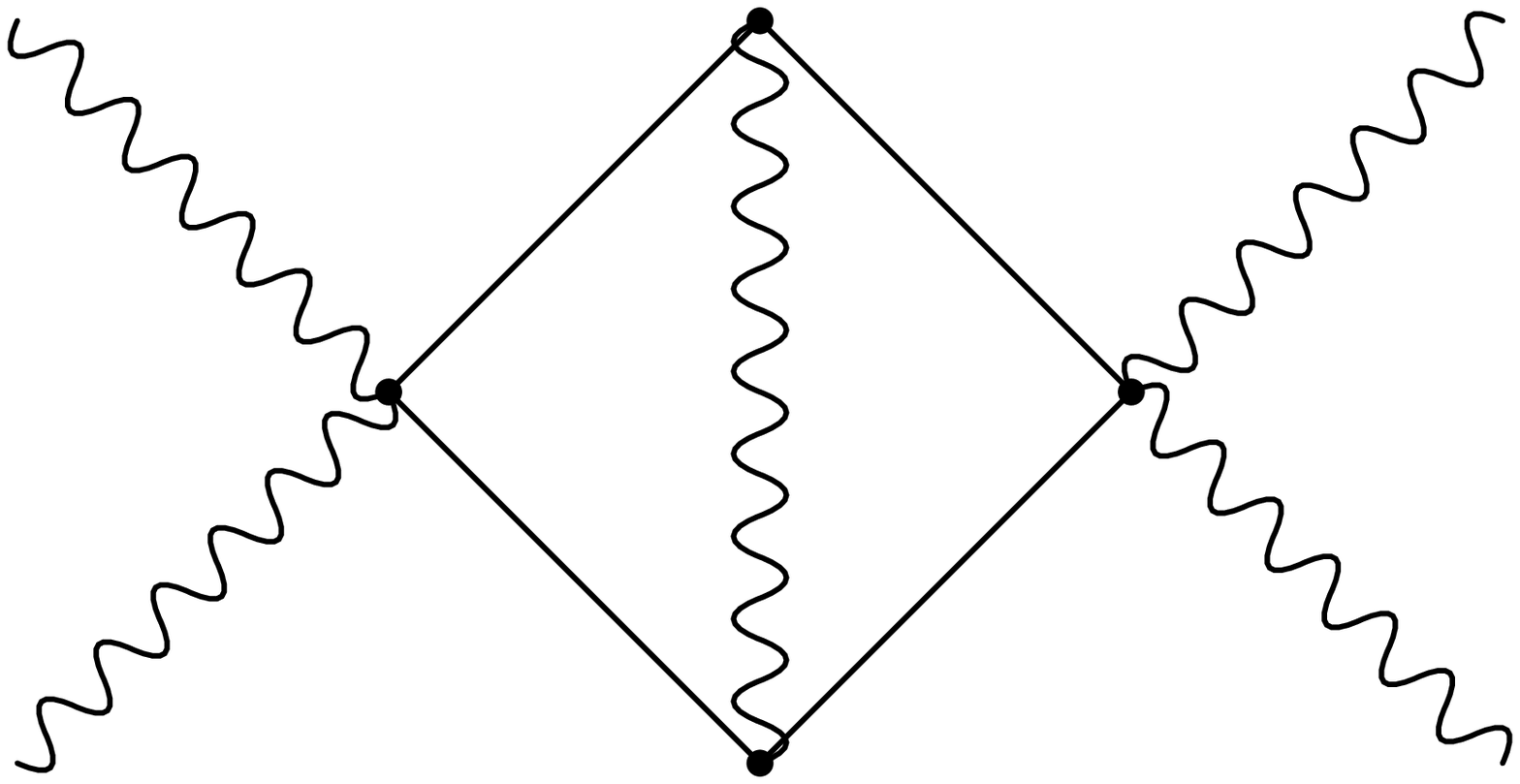}}\\
O&P&Q
\end{tabular}
\caption{Graphs relevant for the scalar electron contribution $\MS{2}{}$. 
The solid lines represent the scalar electron while the wavy lines denote 
the photon.}
}
\clearpage

\FIGURE[t]{
\label{fig:D2}
\begin{tabular}{ccc}
\scalebox{.2}{\includegraphics{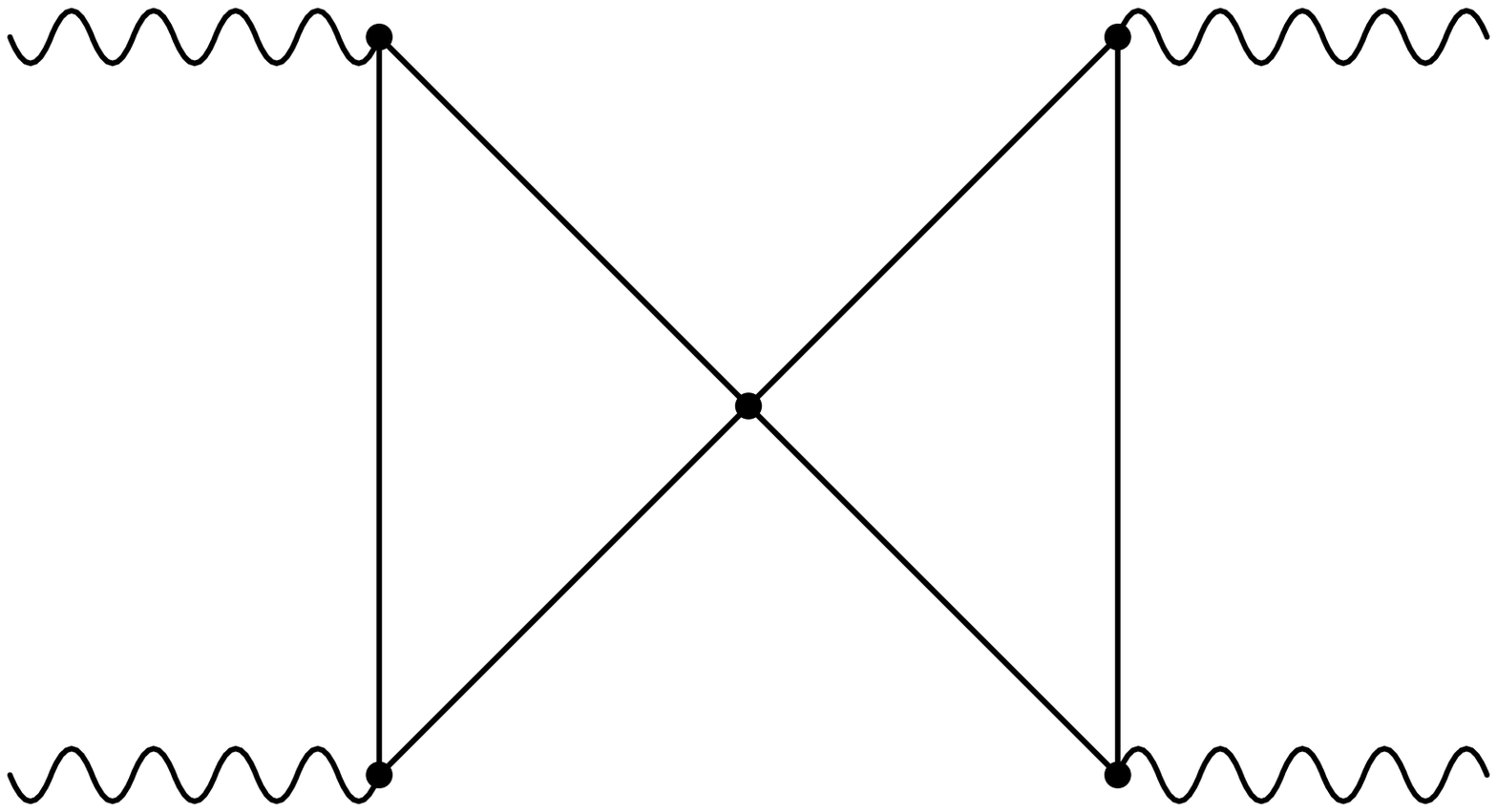}}&
\scalebox{.2}{\includegraphics{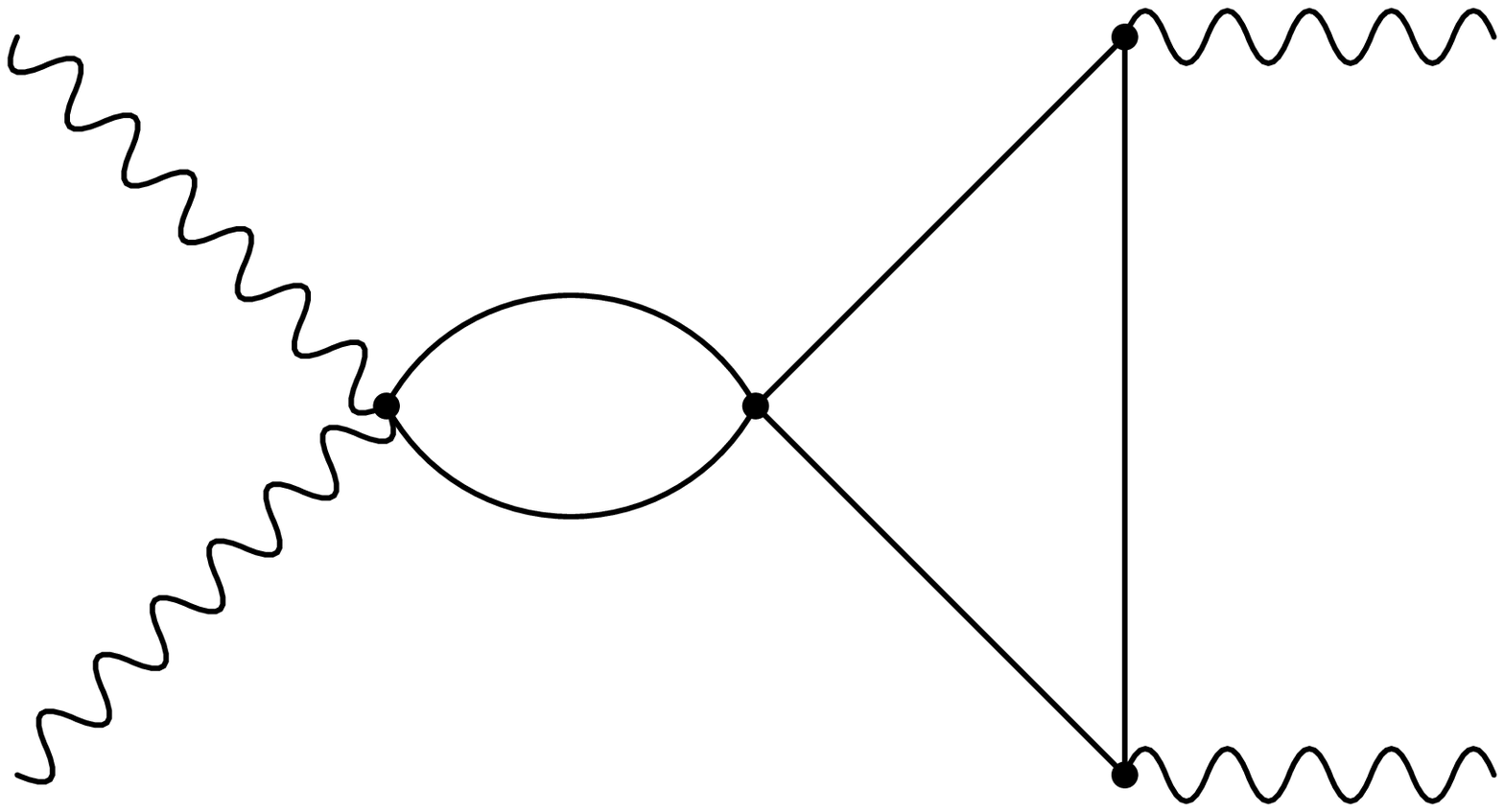}}&
\scalebox{.2}{\includegraphics{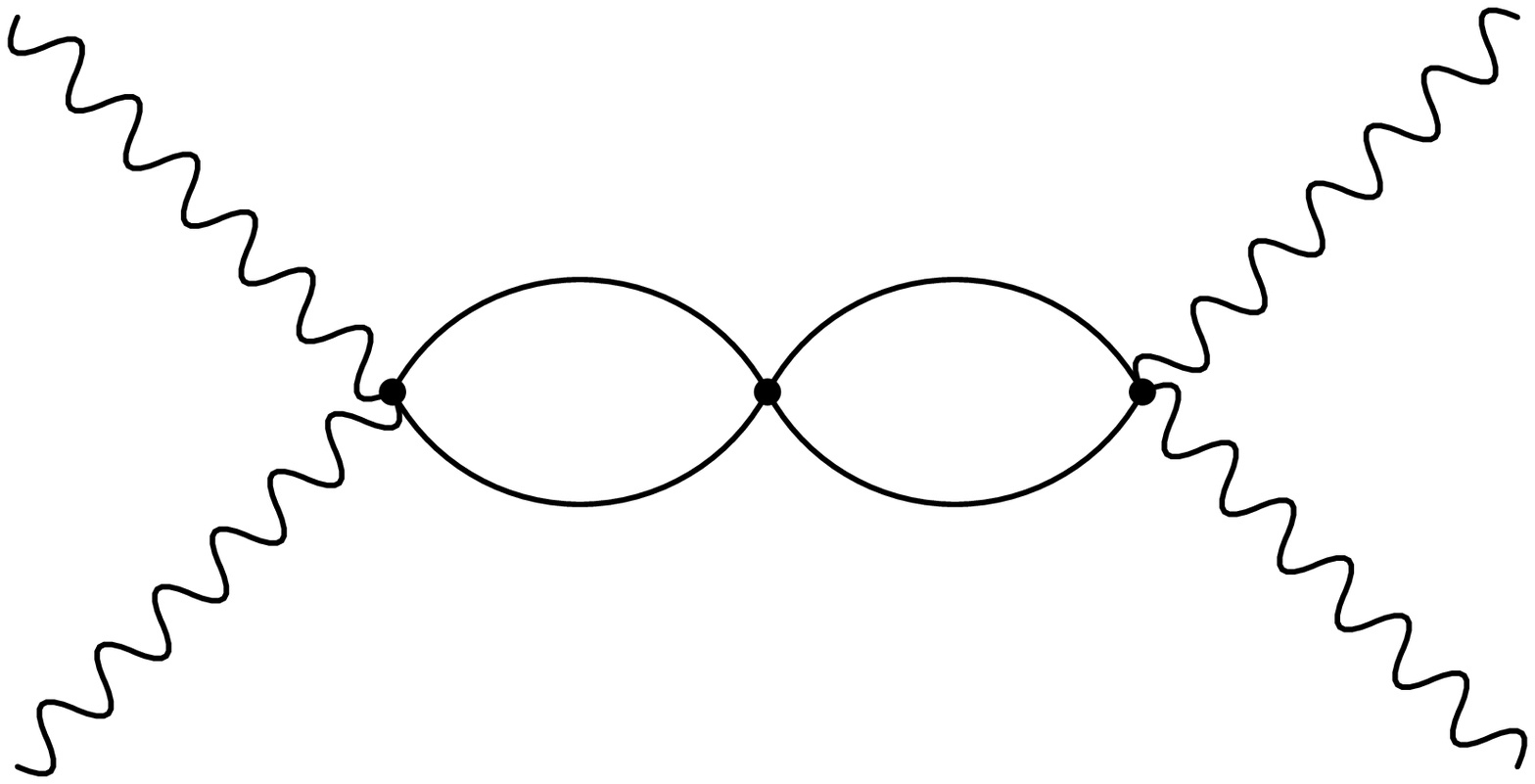}}\\
R&S&T
\end{tabular}
\caption{Graphs relevant for the scalar four-point interaction contribution
 $\MV{2}{}$.
The solid lines represent the scalar electron while the wavy lines denotes
the photon.}
}

\FIGURE[h]{
\label{fig:E2}
\begin{tabular}{ccc}
\scalebox{.2}{\includegraphics{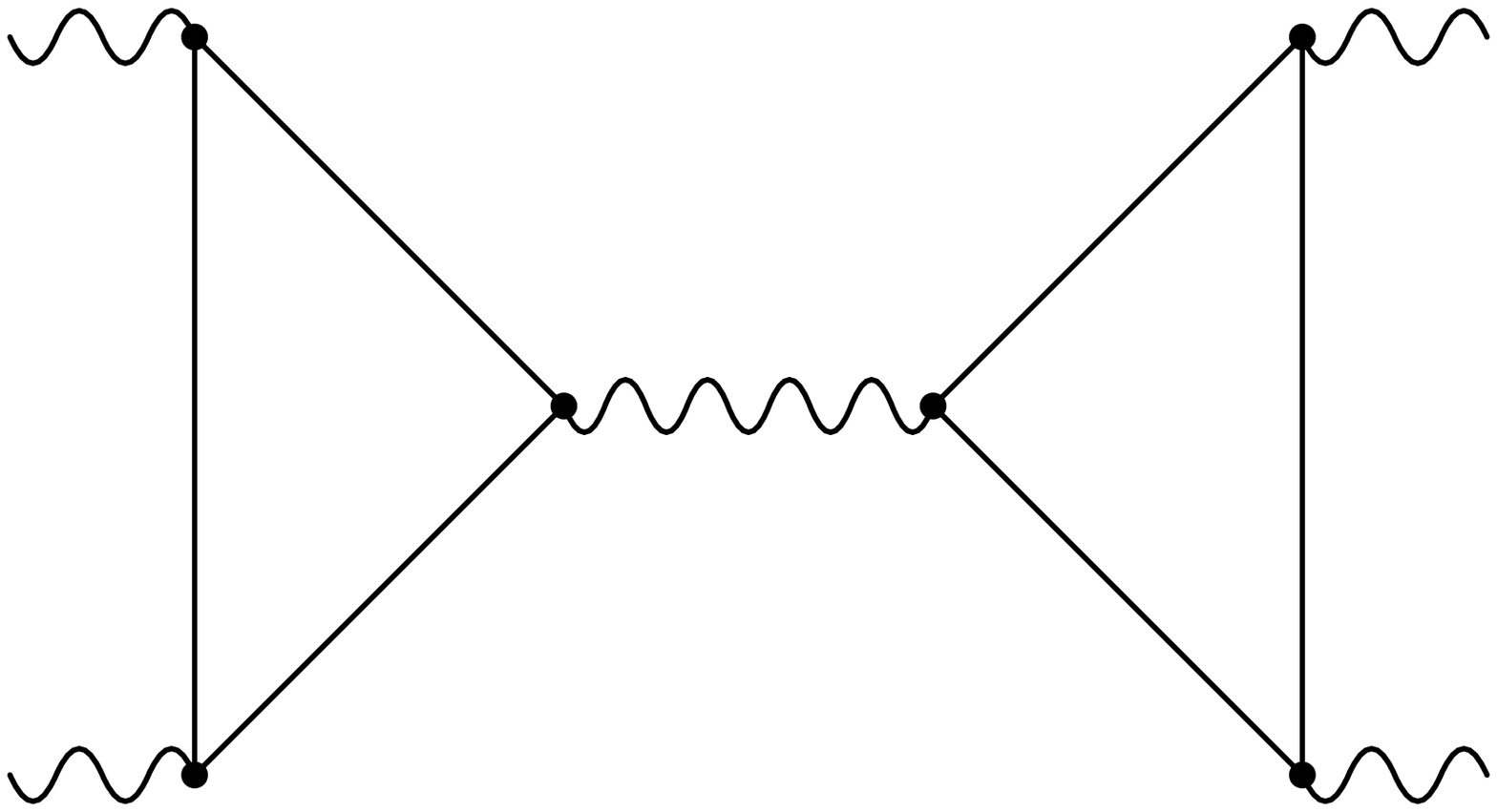}}&
\scalebox{.2}{\includegraphics{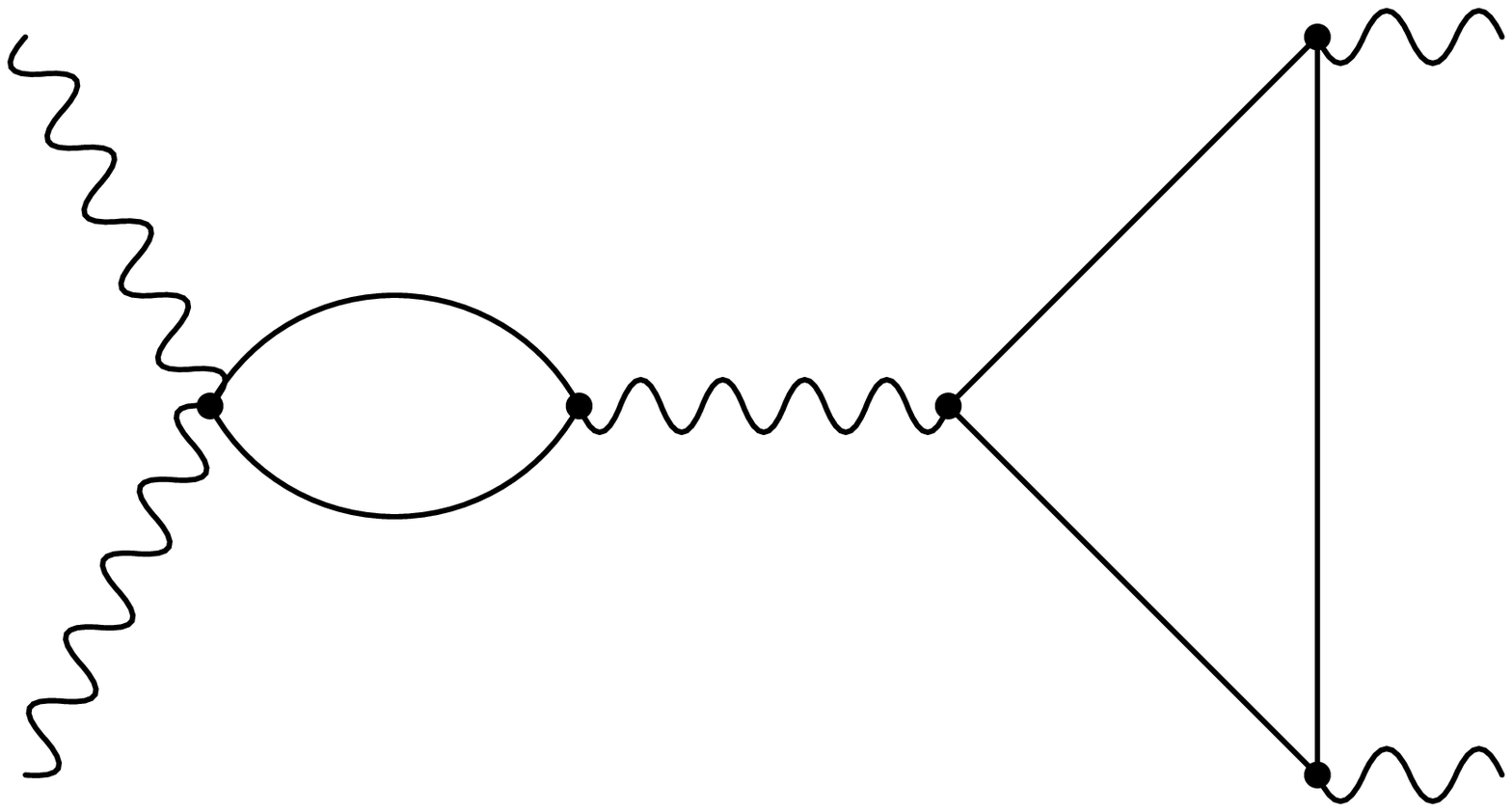}}&
\scalebox{.2}{\includegraphics{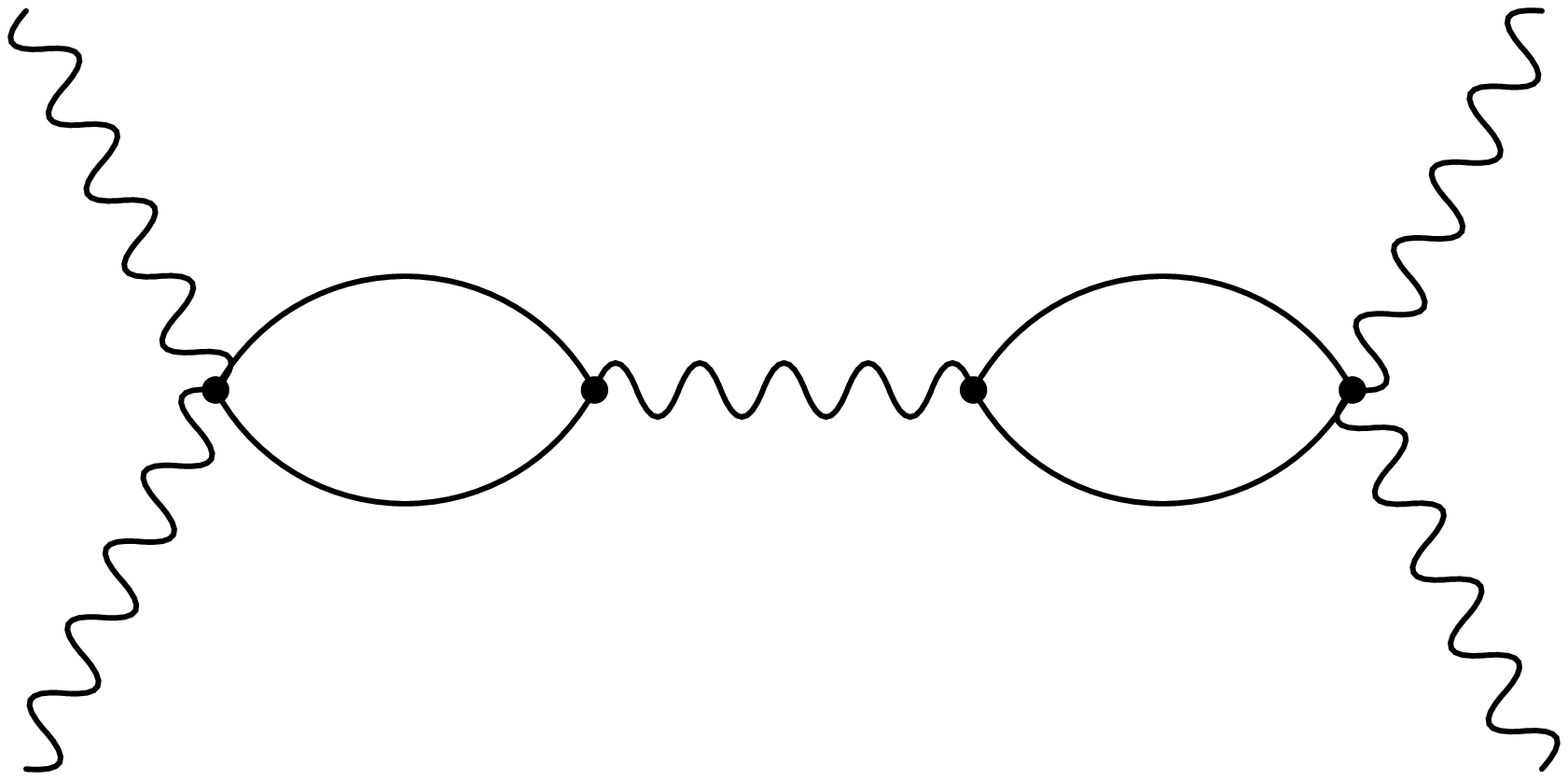}}\\
U&V&W
\end{tabular}
\caption{One-particle reducible graphs relevant for the electron,  
scalar electron and scalar photon contributions $\MF{2}{}$, $\MS{2}{}$ and $\MX{2}{}$. 
The solid lines represent particles from the matter multiplet, i.e. the electron
or scalar electron while the wavy lines denote the photon or scalar photon. Graphs $V$ and $W$ 
are only part of the scalar electron contribution.}
}

Explicit evaluation of the two-loop graphs yields rather lengthy 
results for the individual gauge invariant subsets, and we list them in  
Appendix~\ref{sec:conts}.
Combining the various components according to Eq.~(\ref{eq:N1sum}) gives the
following two-loop $N=1$ SUSY QED helicity amplitudes,
\begin{eqnarray}
\M{2}{++++}&=&0 +{\cal O}(\epsilon),\nonumber \\
\M{2}{+++-}&=&0 +{\cal O}(\epsilon),\nonumber \\
\M{2}{++--}&=&
+{}\Biggl ({}-{16}\,{\bb}\,{\Lidy}+{8}\,{\bb}\,{}\Biggl ({\Ly}+{2}\Biggr ){}\,{\Licy}+{8}\,{\bb}\,{}{\Ly}{}\,{\Licx}
\nonumber \\ &&
+{8}\,{\bb}\,{}\Biggl ({\Lx}\,{\Ly}-{\Ly}+{\Lx}\Biggr
){}\,{\Liby}-{16}\,{\bb}\,{\zeta_3}+{16\over 45}\,{\bb}\,{\pi^4}-{2\over
3}\,{\bb}\,{}\Biggl ({6}\,{\Ly}+{11}+{2}\,{\Lx}\,{\Ly}\Biggr ){}\,{\pi^2}
\nonumber \\ &&
+{2\over 3}\,{\bb}\,{}\Biggl ({}-{\Lx^4}+{6}\,{\Lx^2}\,{\Ly^2}+{4}\,{\Lx}\,{\Ly^3}+{12}\,{\Lx}\,{\Ly}-{4}\,{\Lx^3}+{8}\,{\Ly^3}\Biggr ){}\Biggr ){}\,{\utss}
\nonumber \\ &&
+{}\Biggl ({}-{16}\,{\bb}\,{\Lidz}-{8}\,{\bb}\,{}\Biggl ({\Ly}-{\Lx}\Biggr ){}\,{\Licy}-{8}\,{\bb}\,{}\Biggl ({\Ly}-{\Lx}\Biggr ){}\,{\Licx}-{8}\,{\bb}\,{\Lx}\,{\Ly}\,{\Liby}
\nonumber \\ &&
+{}\Biggl ({}-{8\over 3}\,{\bb}\,{\pi^2}-{8}\,{\bb}\,{\Lx}\,{\Ly}\Biggr ){}\,{\Libx}+{1\over 15}\,{\bb}\,{\pi^4}-{2\over 3}\,{\bb}\,{}\Biggl ({\Ly}+{2}-{\Lx}\Biggr ){}\,{}\Biggl ({\Ly}-{\Lx}\Biggr ){}\,{\pi^2}
\nonumber \\ &&
+{1\over 3}\,{\bb}\,{}\Biggl ({4}\,{\Lx^3}+{12}\,{\Lx^3}\,{\Ly}+{12}\,{\Lx}\,{\Ly^2}-{12}\,{\Lx^2}\,{\Ly}-{4}\,{\Lx}\,{\Ly^3}+{\Ly^4}-{3}\,{\Lx^4}
\nonumber \\ &&
-{30}\,{\Lx^2}\,{\Ly^2}-{4}\,{\Ly^3}\Biggr ){}\Biggr ){}\,{\tttu } 
+{ 8}\,{\bb}\,{\Lx^2}\,{\sstt } 
\nonumber \\ &&
+{i\pi }\Biggl\{{ 16}\,{\bb}\,{\Lx}\,{\sstt } 
+{}\Biggl ({4\over 3}\,{\bb}\,{}\left ({\Ly}-{\Lx}\right){}\,{\pi^2}+{4\over
3}\,{\bb}\,{}\left ({\Ly}-{\Lx}\right){^3}\Biggr ){}\,{\tttu } \nonumber \\ &&
\nonumber \\ &&
+{}\Biggl ({16}\,{\bb}\,{\Licx}+{8}\,{\bb}\,{\Lx}\,{\Liby}+{8}\,{\bb}\,{\Lx}\,{\Libx}
\nonumber \\ &&
-{4\over 3}\,{\bb}\,{}\Biggl ({2}\,{\Ly}-{1}\Biggr ){}\,{\pi^2}-{4\over
3}\,{\bb}\,{\Ly}\,{}\Biggl({6}\,{\Lx}-{6}\,{\Ly}+{\Ly^2}-{9}\,{\Lx}\,{\Ly}-{12}\Biggr
){}\Biggr ){}\,{\utss} \Biggr \}\nonumber \\
&&
+ \Biggl \{ u \leftrightarrow t \Biggr \} + {\cal O}(\epsilon) , 
\label{eq:N1result}
\end{eqnarray}
where we have used the standard polylogarithm identities~\cite{kolbig} to
express 
our results in terms of a basis set of constants
(where $\zeta_n$ is the Riemann Zeta function, $\zeta_2 = \pi^2/6$, 
$\zeta_3 = 1.202056\ldots$),   
logarithms and polylogarithms 
${\rm Li}_n(w)$ defined by
\begin{eqnarray}
 {\rm Li}_n(w) &=& \int_0^w \frac{dt}{t} {\rm Li}_{n-1}(t) \qquad {\rm ~for~}
 n=2,3,4\nonumber \\
 {\rm Li}_2(w) &=& -\int_0^w \frac{dt}{t} \log(1-t).
\label{eq:lidef}
\end{eqnarray} 
with arguments $x$, $1-x$ and $(x-1)/x$, where
\begin{equation}
\label{eq:xydef}
x = -\frac{t}{s}, \qquad y = -\frac{u}{s} = 1-x, 
\qquad z=-\frac{u}{t} = \frac{x-1}{x}.
\end{equation}
In the physical region $s>0$ and $t,u<0$, our basis set of functions are all
real. 

As expected from the SUSY Ward identities, Eqs.~(\ref{eq:swipppp}) and 
(\ref{eq:swipppm}), two of the helicity amplitudes vanish in the $n \to 4$
limit.   However, the identities are violated by terms of ${\cal
O}(\epsilon)$ which can be traced back to the SUSY breaking nature of
dimensional regularisation.

The SUSY Ward identity does not require that ${\cal M}_{++--}$ vanishes, but,
as in the one-loop case, there are still significant cancellations and the full
amplitude is somewhat more compact than the individual contributions (which are
listed in Appendix~\ref{sec:conts}): In fact, at two-loops we expect that
amplitudes contain  terms up to weight 4 (counting ${\rm Li}_n$, $\ln^n$ and
$\zeta_n$ as weight $n$, $\ln^m {\rm Li}_n$ as weight $n+m$ and so on).  Each
of the  individual contributions for ${\cal M}_{++--}$   listed in
Appendix~\ref{sec:conts}, and in particular the fermion contribution
$\MF{2}{++--}$ that corresponds to the Standard Model ($N=0$ SUSY),
demonstrates that all possible weights,  $ 0,\ldots,4$,  are present.  
However, we see that the $N=1$ SUSY amplitude of Eq.~(\ref{eq:N1result})
contains terms of weight 2, 3 and 4 and that,  as in the one-loop case, terms
of weight 0 and 1 cancel. We also note that the individual contributions
contain terms proportional to the dimensionless ratios, $t^4/u^2s^2$,
$t^2/s^2$, $t/u$  (and $ t \leftrightarrow u$), while in the combination
selected by $N=1$ SUSY, (\ref{eq:N1result}), the terms proportional to $t^4/u^2s^2$ drop out.

\subsubsection{$N=2$ SUSY QED}

For $N=2$ SUSY, there are additional contributions from the gaugino, the scalar photon and
the modified quartic scalar electron interactions.   
In terms of the different gauge-invariant pieces we find that
\begin{equation}
\label{eq:N2sum}
\M{2}{}
=\MS{2}{}
+\MF{2}{}
+2\MP{2}{}
+3\MV{2}{}
+\MX{2}{},
\end{equation}
where again the dependence on the helicities has been suppressed.
Here, $\MX{2}{}$ denotes the 144 graphs involving the scalar photon while the factors of 
2 and 3 multiplying the photino and four-point scalar electron contributions 
respectively reflect the 
aditional photino and the modified scalar electron interactions of the
$N=2$ SUSY theory. 

Combining the individual gauge invariant contributions listed in
Appendix~\ref{sec:conts} according to Eq.~(\ref{eq:N2sum}), 
we find the $N=2$ SUSY QED helicity amplitudes are rather compact and are given by,
\begin{eqnarray}
\M{2}{++++}&=&0 +{\cal O}(\epsilon),\nonumber \\
\M{2}{+++-}&=&0 +{\cal O}(\epsilon),\nonumber \\
\M{2}{++--}&=&
\Biggl ({}-{16}\,{\bb}\,{\Lidy}+{8}\,{\bb}\,{\Ly}\,{\Licx}+{8}\,{\bb}\,{\Ly}\,{\Licy}+{16\over 45}\,{\bb}\,{\pi^4}-{2\over 3}\,{\bb}\,{\Lx}\,{\Ly}\,{\pi^2}
\nonumber \\ &&
-{2\over 3}\,{\bb}\,{\Ly^3}\,{}\Biggl ({\Ly}-{4}\,{\Lx}\Biggr ){}\Biggr ){}\,{\utss}\nonumber \\
&& +{i\pi }\Biggl\{\Biggl ({16}\,{\bb}\,{\Licx}-{4\over 3}\,{\bb}\,{\Ly}\,{\pi^2}-{4\over 3}\,{\bb}\,{\Ly^2}\,{}\Biggl
({\Ly}-{3}\,{\Lx}\Biggr ){}\Biggr ){}\,{\utss}\Biggr \}  
\nonumber \\
&&
+ \Biggl \{ u \leftrightarrow t \Biggr \} + {\cal O}(\epsilon). 
\label{eq:N2result}
\end{eqnarray}
As expected, two of the helicity amplitudes vanish as $n \to 4$ due to the SUSY
Ward identity. The remaning
non-trivial helicity amplitude is considerably simpler than that obtained in
either pure
QED~\cite{Bern:2001dg} (see the fermion loop contributions in
Appendix~\ref{sec:conts})
or the $N=1$ SUSY QED helicity amplitudes of Eq.~(\ref{eq:N1result}).
In particular, we note that only terms of weight 4 remain, the contributions of
weight 2 and 3 (that were present in the
$N=1$ case) have cancelled.    Furthermore, all terms depending on the ratios of
kinematic scales have dropped out. 

\newpage
\section{Summary}
\label{sec:summary}

We have demonstrated that the method based on $n$-dimensional projections is
able to generate helicity amplitudes in an efficient way. We have been able to
confirm the previous results for photon-photon scattering via a charged fermion
loop obtained by helicity methods in the high energy limit where the fermion
mass can be ignored. The method can, in principle, be used for more complicated
processes, involving massive particles in the loop,  non-abelian fields and/or
more external vertices. The method is constructed such that one can use
standard $n$-dimensional Lorentz covariant reduction techniques for
tensor-integrals not only for squared matrix elements, but also for helicity
amplitudes.

As an application we studied photon-photon scattering in the theoretically
interesting cases of $N=1$ and $N=2$ supersymmetric QED. Because the process is
both ultraviolet and infrared convergent, the results are explicitly
supersymmetric in $n=4$, which is sufficient to draw conclusions regarding SUSY
cancellations. We have not addressed the question of whether the method can be
used to define supersymmetric amplitudes also at higher order in $(n-4)$.
However, we expect that the situation here will be similar to that found  using
other techniques, where explicit terms proportional to $(n-4)$  have to be
added in order to preserve the SUSY Ward identities at all orders in $(n-4)$. 

The results we found have an interesting pattern of simplification as one
increases the number of supersymmetries. In normal QED, corresponding to $N=0$
supersymmetries, one finds three independent amplitudes.    For supersymmetric
theories, two of these helicity amplitudes vanish.  This is well understood on
the basis of the supersymmetric  Ward identity. An intricate pattern of
cancellations occurs in the remaining (non-trivial) amplitude, ${\cal
M}_{++--}$, as $N$ is
increased, both in terms of the dimensionless ratios of kinematic scales and in
the weights of the functions present in the amplitude. 

\TABLE[b]{
\begin{tabular}{|c|c|c|}\hline
\qquad ${\cal M}_{++--}$    \qquad & \qquad   one-loop         \qquad & \qquad   two-loop             \qquad \\ \hline
\qquad   $N=0$  \qquad & \qquad   0,1,2            \qquad & \qquad   0,1,2,3,4            \qquad \\
\qquad   $N=1$  \qquad & \qquad   \phantom{0,1,}2  \qquad & \qquad   \phantom{0,1,}2,3,4  \qquad \\
\qquad   $N=2$  \qquad & \qquad   \phantom{0,1,}2  \qquad & \qquad   \phantom{0,1,2,3,}4  \qquad \\ \hline
 \end{tabular}
\label{tab:weights}
\caption{Weights of terms contribution to ${\cal M}_{++--}$ at one- and two-loops.   The $N=0$ result refers to
the fermion contribution of Eq.~(\ref{eq:fermion}) while values given for  $N=1$ and $N=2$ are extracted from
Eqs.~(\ref{eq:N1result}) and (\ref{eq:N2result}). }}

\TABLE[t]{
\begin{tabular}{|c|c|c|c|}\hline
& & & \\
 ${\cal M}_{++--}$     &    \qquad 1 \hspace{1cm} &  
${t^2\over s^2},~{u^2\over s^2},~{t\over u},~{u\over t}$  &  ${t^4\over u^2s^2},~{u^4\over t^2s^2}$  \\ 
& & & \\\hline
   $N=0$   &   $\sqrt{}$    &    $\sqrt{}$       &   $\sqrt{}$   \\
   $N=1$   &   $\sqrt{}$   &    $\sqrt{}$  & \\
   $N=2$   &   $\sqrt{}$   &	   & \\ \hline
 \end{tabular}
\label{tab:ratios}
\caption{Ratios of kinematic scales appearing in the two-loop amplitude ${\cal M}_{++--}$
for different amounts of supersymmetry.}
}

First, considering the one-loop level, one finds in normal ($N=0$) QED terms
proportional to the dimensionless ratios $t^2/s^2$ and $u^2/s^2$ together with
weights of logarithm of 0,1,~2 (see Eq.~(\ref{eq:oneconts}). In the
supersymmetric case (\ref{eq:oneresult}), 
there are no dimensionless ratios and there is a uniform weight of 2 for
the logarithms. 

At the two-loop level, the pattern is similar, see Tables~\ref{tab:weights} and
\ref{tab:ratios}. Normal ($N=0$)  QED contains powers of $t/s$ up to 4 and all
weights of logarithm from 0 to 4. The $N=1$ case shows some simplifications,
powers of $t/s$ exist only up to squares and the logarithmic terms have weights
2, 3, 4. Finally in the $N=2$ theory there are no powers of $t/s$ in the 
amplitudes and the logarithms are uniformly of weight 4. We see therefore, that
increasing the number of supersymmetries reduces the complexity of the
amplitudes and that the $N=2$ theory is maximally simplified.  A conjecture
would be that this pattern persists in higher orders of perturbation theory,
where one would expect no powers and a weight of two times the level of loops
in the logarithms.  This pattern might be explainable by performing similar
calculations in superspace.

\section*{Acknowledgements}
We thank Valya Khoze, Adrian Signer, Bas Tausk and Georg Weiglein for helpful
discussions. This work was supported in part by the UK Particle Physics and
Astronomy  Research Council,  by the EU Fifth Framework Programme `Improving
Human Potential', Research Training Network `Particle Physics Phenomenology  at
High Energy Colliders', contract HPRN-CT-2000-00149 and by the
DFG-Forschergruppe Quantenfeldtheorie, Computeralgebra und Monte-Carlo
Simulation. We thank the British Council and German Academic Exchange Service
for support under ARC project 1050.  PM acknowledges the support of 
the German Academic Exchange Service.

\newpage
\appendix
\section{The SUSY QED Feynman Rules}
\label{sec:rules}
The Feynman rules for the SUSY QED Lagrangians of Eqs.~(\ref{LSQED1}) and
~(\ref{LSQED2}) with all
momenta incoming are given
by,\vspace{-0.5cm}
\begin{center}
\begin{equation}
\begin{array}{ll}
\vspace{0.5cm}
\parbox{4cm}{\scalebox{.2}{\includegraphics{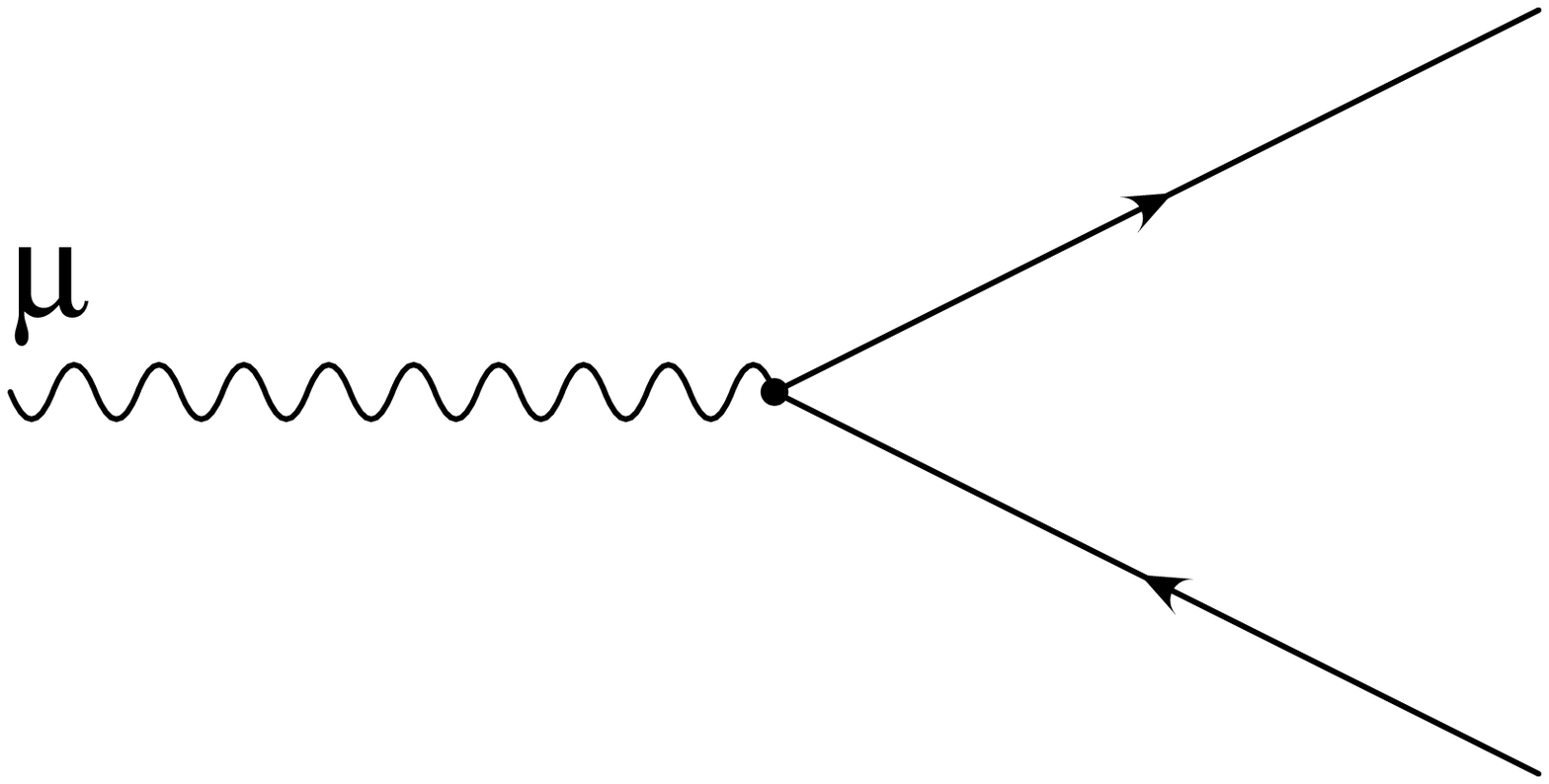}}}&  -ie \gamma^\mu \\
\vspace{0.5cm}
\parbox{4cm}{\scalebox{.2}{\includegraphics{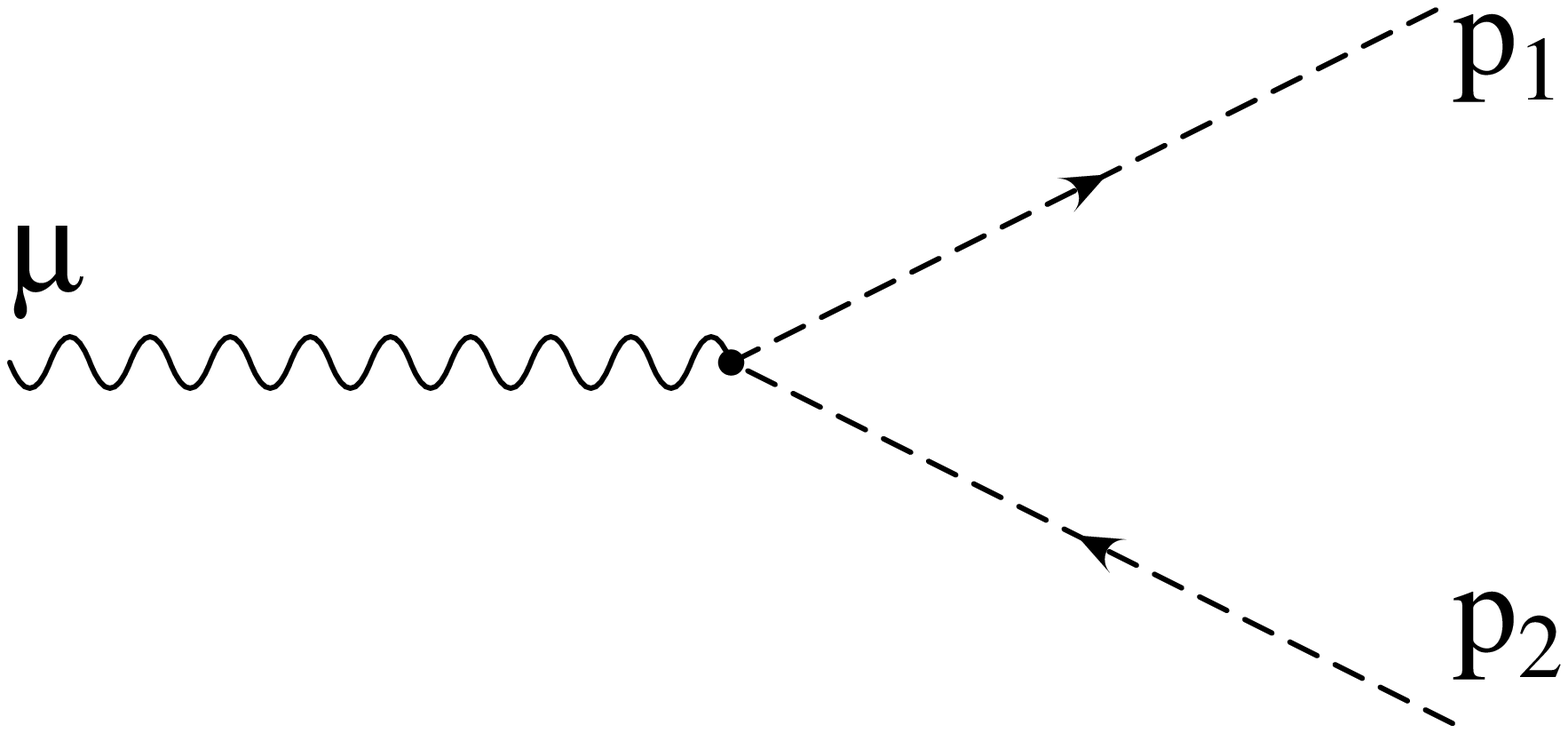}}}&  ie (p_1^\mu-p_2^\mu)\\
\vspace{0.5cm}
\parbox{4cm}{\scalebox{.2}{\includegraphics{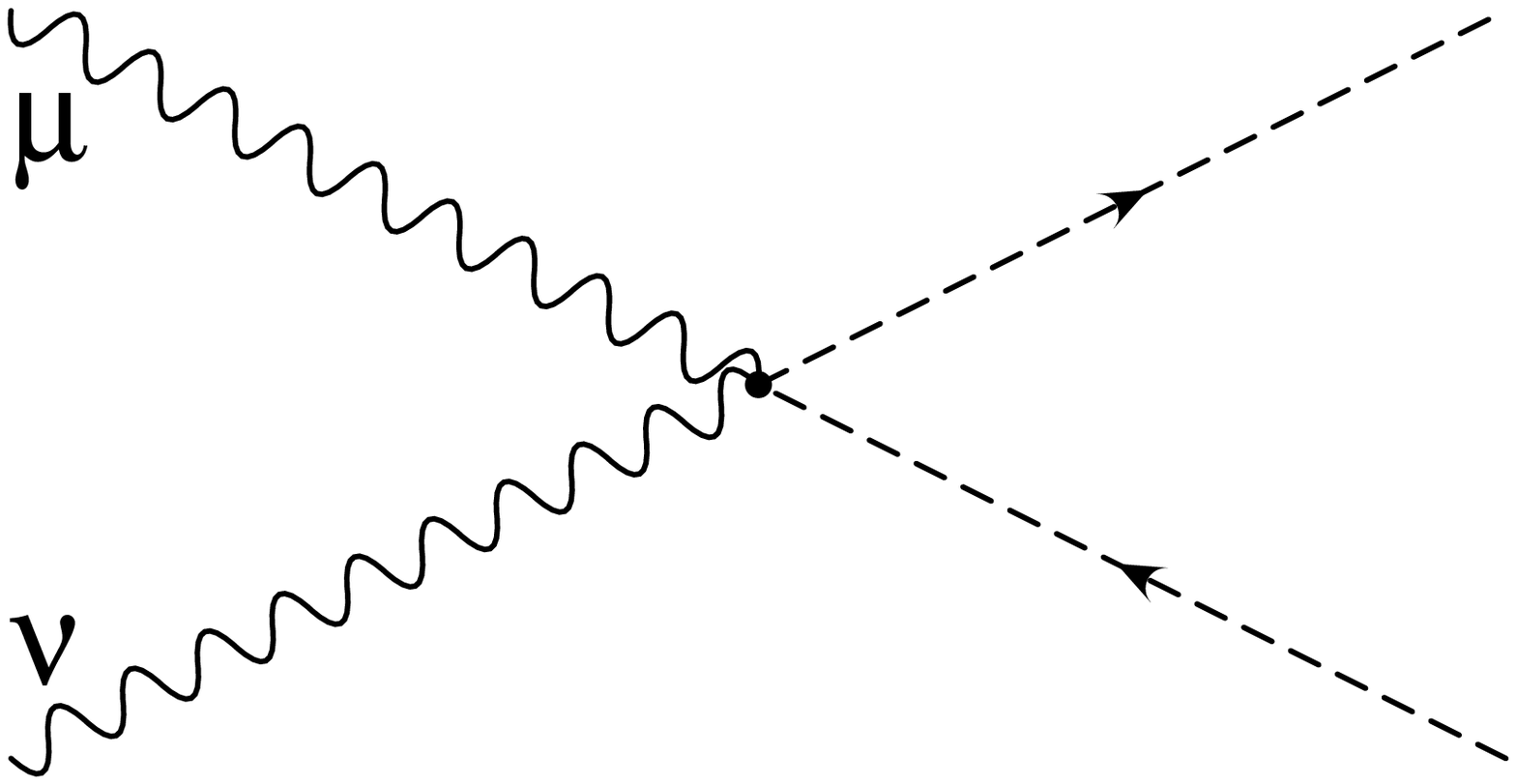}}}&  2ie^2g^{\mu\nu}\\
\vspace{0.5cm}
\parbox{4cm}{\scalebox{.2}{\includegraphics{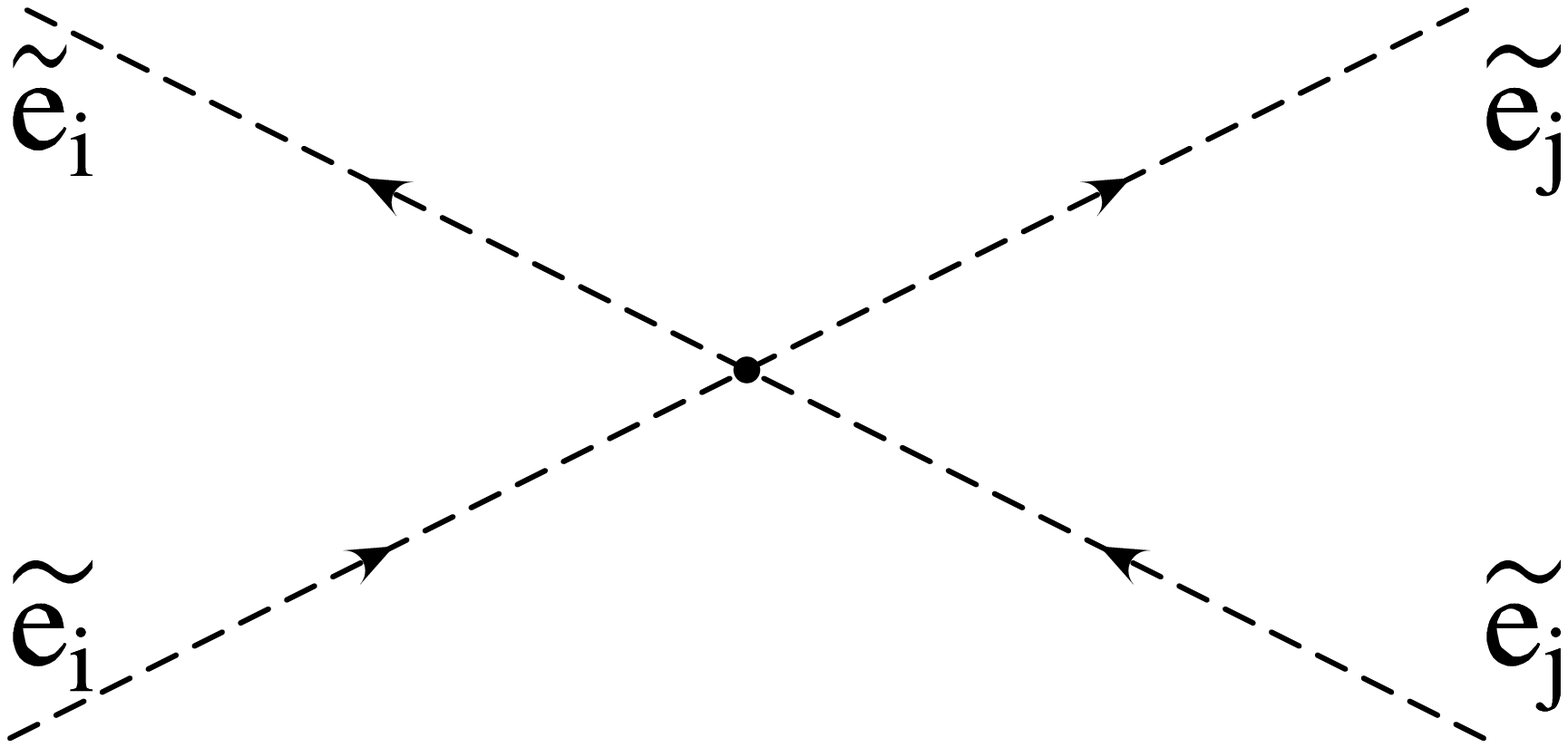}}}& 
\left\{\begin{array}{lc}2ie^2\qquad i=j &\\ 
-ie^2\qquad i\neq j, & N=1 \\
+ie^2\qquad i\neq j, & N=2 \\
\end{array}\right.\\
\vspace{0.5cm}
\parbox{4cm}{\scalebox{.2}{\includegraphics{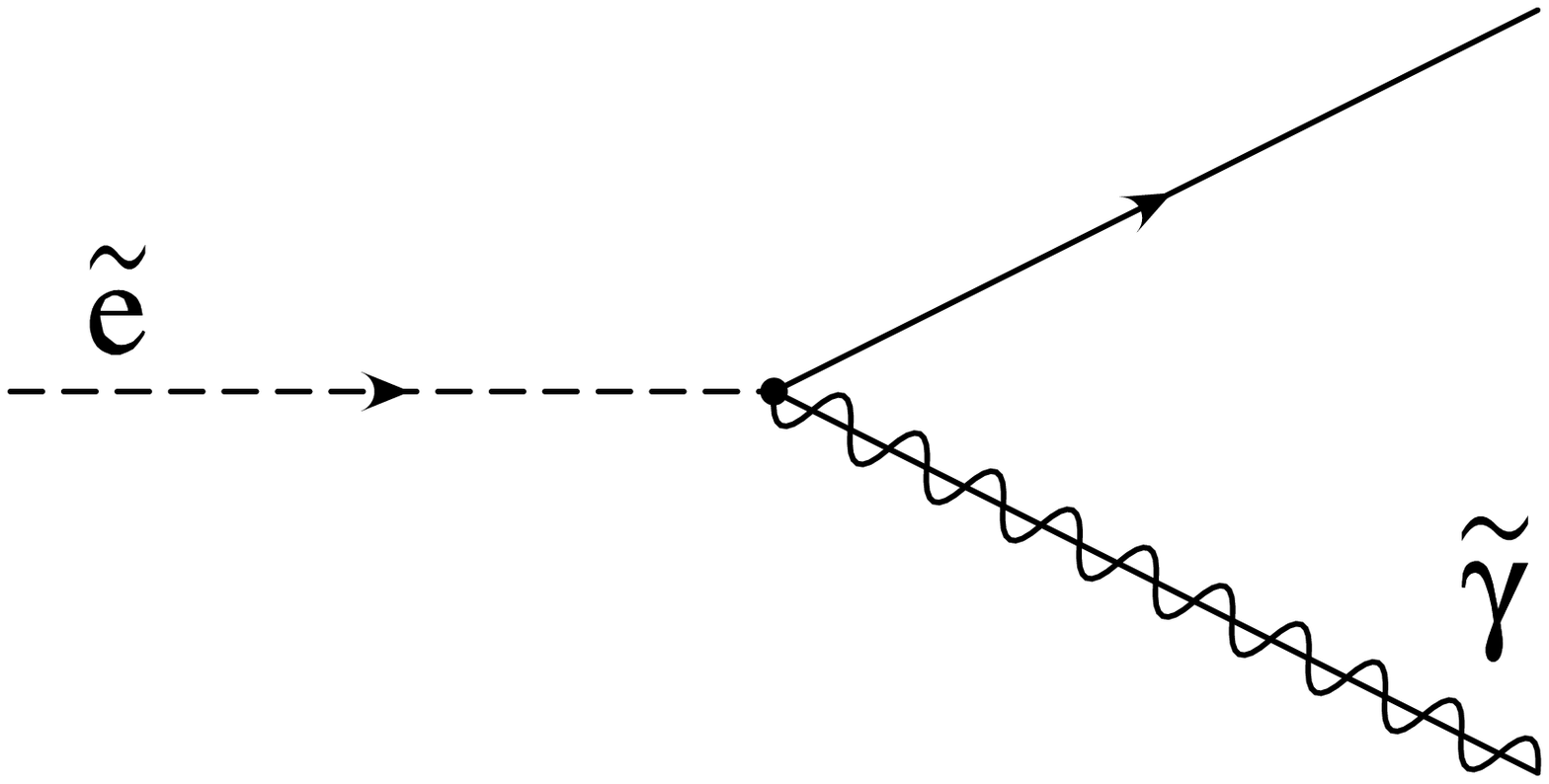}}}&  
\left\{ \begin{array}{l c} 
-ie \sqrt{2}P_R\qquad \tilde{e}=\phi_L^-, & \tilde\gamma=\tilde\gamma_1\\
-ie \sqrt{2}P_L\qquad \tilde{e}=\phi_L^-, & \tilde\gamma=\tilde\gamma_2\\
+ ie\sqrt{2}P_L\qquad \tilde{e}=\phi_R^{+\dag}, &\tilde\gamma = \tilde\gamma_1\\
- ie\sqrt{2}P_R\qquad \tilde{e}=\phi_R^{+\dag}, &\tilde\gamma = \tilde\gamma_2 
\end{array}\right. \\
\vspace{0.5cm}
\parbox{4cm}{\scalebox{.2}{\includegraphics{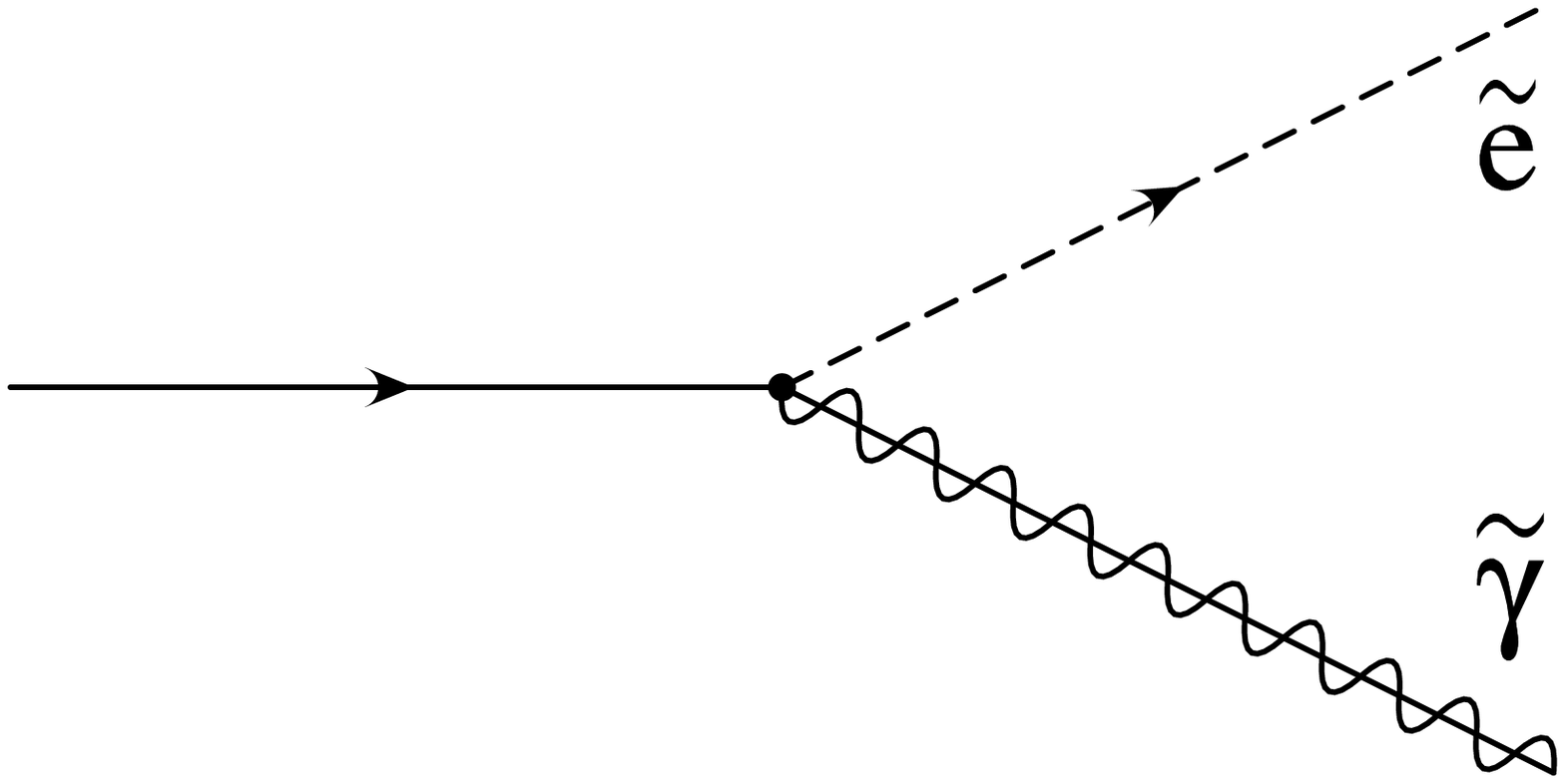}}}&
\left\{ \begin{array}{l c} 
-ie \sqrt{2}P_L\qquad \tilde{e}=\phi_L^{-\dag}, & \tilde\gamma=\tilde\gamma_1\\
-ie \sqrt{2}P_R\qquad \tilde{e}=\phi_L^{-\dag}, & \tilde\gamma=\tilde\gamma_2\\
+ ie\sqrt{2}P_R\qquad \tilde{e}=\phi_R^{+}, &\tilde\gamma = \tilde\gamma_1\\
- ie\sqrt{2}P_L\qquad \tilde{e}=\phi_R^{+}, &\tilde\gamma = \tilde\gamma_2 
\end{array}\right. \\
\vspace{0.5cm}
\parbox{4cm}{\scalebox{.2}{\includegraphics{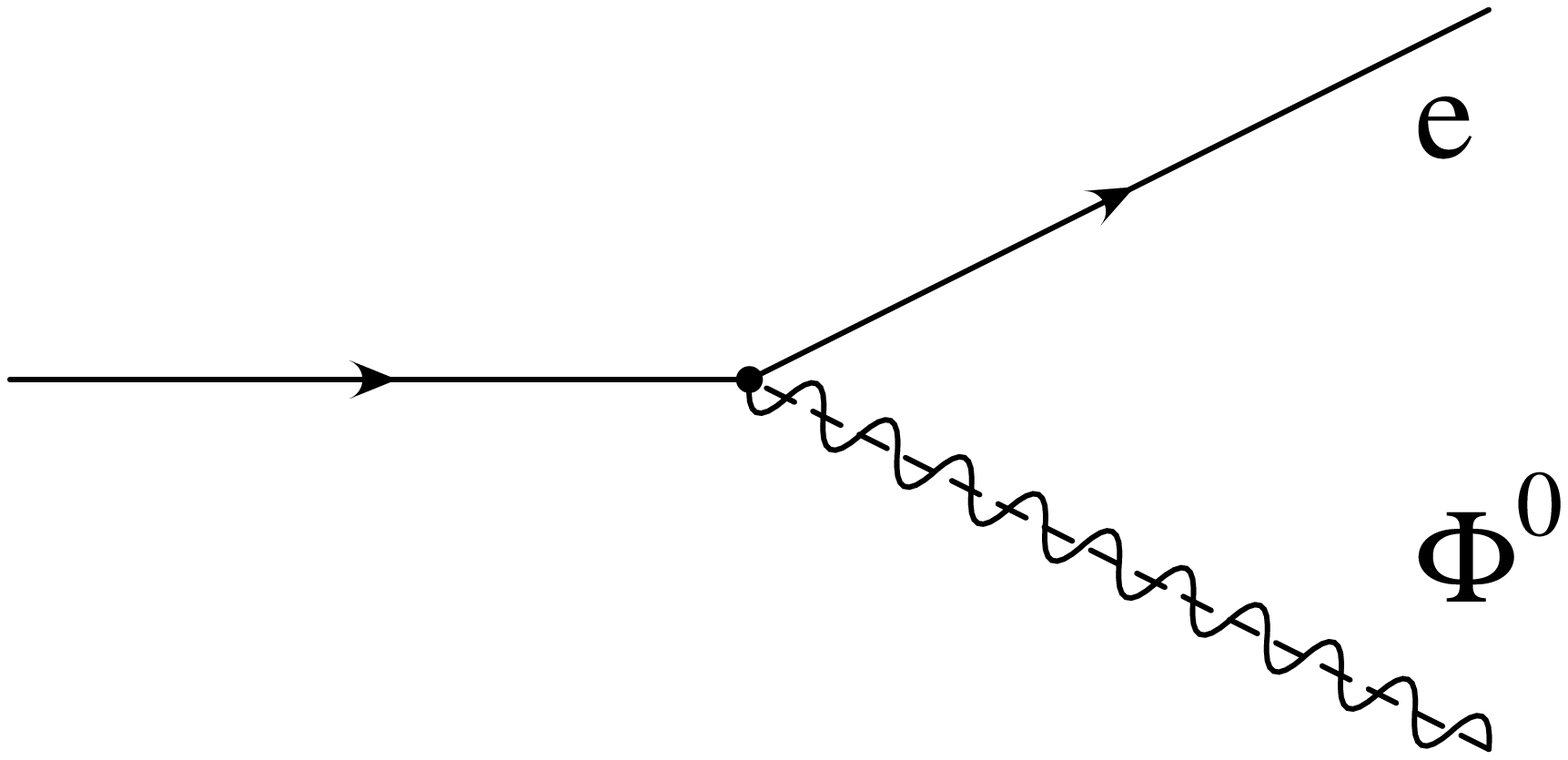}}}&\left\{ \begin{array}{l}
-ie \sqrt{2}P_L\qquad \Phi^0=\phi^0 \\-ie\sqrt{2}P_R\qquad \Phi^0=\phi^{0*} \end{array}\right . 
\end{array}
\end{equation}
\end{center}
where the photon, electron, scalar electron, photino and scalar photon
are denoted by
wavy, solid, dashed, overlaid solid and wavy and 
overlaid dashed and wavy lines respectively.

\newpage

\section{Two-loop contributions to the helicity amplitudes}
\label{sec:conts}

In this appendix we list the individual contributions to the two-loop
helicity amplitudes.   There are five separate gauge invarariant contributions,
 loops with scalar particles coupling to photons, electron loops, 
 photino exchange graphs,  diagrams with the four scalar vertex and graphs
 involving the scalar photon.
We find that individual two-loop contributions are given by,
\begin{eqnarray}
\MS{2}{++++} &=& -6\nonumber \\
\MF{2}{++++} &=& -12\nonumber \\
\MP{2}{++++} &=& +24\nonumber \\
\MV{2}{++++} &=& -6 \nonumber \\
\MX{2}{++++} &=& -12 \\
\MS{2}{+++-}&=&{ }-{4}\,{\bb}\,{\Lx^2}\,{\tfiveou } +{}\Biggl ({\bb}\,{\pi^2}-{2}\,{\bb}\,{}\Biggl ({2}\,{\Ly}+{\Lx}\,{\Ly}-{3}\,{\Lx^2}\Biggr ){}\Biggr ){}\,{\utss}-{4}\,{\bb}\,{\Lx}\,{}\Biggl ({2}+{3}\,{\Lx}\Biggr ){}\,{\sstt } 
\nonumber \\ &&
+{}\Biggl ({}-{2}\,{\bb}\,{\pi^2}-{2}\,{\bb}\,{}\Biggl ({2}\,{\Ly}-{2}\,{\Lx}-{2}\,{\Lx}\,{\Ly}-{3}\,{\Lx^2}+{3}\,{\Ly^2}\Biggr ){}\Biggr ){}\,{\tttu } 
\nonumber \\ &&
+{i\pi }\Biggl\{-{8}\,{\bb}\,{}\Biggl ({1}+{3}\,{\Lx}\Biggr ){}\,{\sstt } +{4}\,{\bb}\,{}\Biggl ({}-{1}+{2}\,{\Lx}\Biggr ){}\,{\utss}-{8}\,{\bb}\,{}\Biggl ({}-{2}\,{\Lx}+{\Ly}\Biggr ){}\,{\tttu } 
-{8}\,{\bb}\,{\Lx}\,{\tfiveou } \Biggr \} \nonumber \\ &&
+ \Biggl \{ u \leftrightarrow t \Biggr \} \nonumber \\  
%%%
\MF{2}{+++-}&=&{ }-{2}\,{\bb}\,{\Lx^2}\,{\tfiveou } -{\bb}\,{}\Biggl ({}-{\Lx^2}+{2}\,{\Ly}\Biggr ){}\,{\utss}-{2}\,{\bb}\,{\Lx}\,{}\Biggl ({2}+{3}\,{\Lx}\Biggr ){}\,{\sstt } 
\nonumber \\ &&
+{}\Biggl ({}-{\bb}\,{\pi^2}-{\bb}\,{}\Biggl ({2}\,{\Ly}-{2}\,{\Lx}-{2}\,{\Lx}\,{\Ly}-{3}\,{\Lx^2}+{3}\,{\Ly^2}\Biggr ){}\Biggr ){}\,{\tttu } 
\nonumber \\ &&
+{i\pi }\Biggl\{-{4}\,{\bb}\,{}\Biggl ({1}+{3}\,{\Lx}\Biggr ){}\,{\sstt } +{2}\,{\bb}\,{}\Biggl ({\Lx}-{1}\Biggr){}\,{\utss}-{4}\,{\bb}\,{}\Biggl ({}-{2}\,{\Lx}+{\Ly}\Biggr ){}\,{\tttu } -{4}\,{\bb}\,{\Lx}\,{\tfiveou } \Biggr \}
\nonumber \\ &&
+ \Biggl \{ u \leftrightarrow t \Biggr \} \nonumber \\  
%%%
\MP{2}{+++-}&=&{ 6}\,{\bb}\,{\Lx^2}\,{\tfiveou } +{}\Biggl ({}-{\bb}\,{\pi^2}+{\bb}\,{}\Biggl ({6}\,{\Ly}-{7}\,{\Lx^2}+{2}\,{\Lx}\,{\Ly}\Biggr ){}\Biggr ){}\,{\utss}+{6}\,{\bb}\,{\Lx}\,{}\Biggl ({2}+{3}\,{\Lx}\Biggr ){}\,{\sstt } 
\nonumber \\ &&
+{}\Biggl ({3}\,{\bb}\,{\pi^2}+{3}\,{\bb}\,{}\Biggl ({2}\,{\Ly}-{2}\,{\Lx}-{2}\,{\Lx}\,{\Ly}-{3}\,{\Lx^2}+{3}\,{\Ly^2}\Biggr ){}\Biggr ){}\,{\tttu } 
\nonumber \\ &&
+{i\pi }\Biggl\{{ 12}\,{\bb}\,{}\Biggl ({1}+{3}\,{\Lx}\Biggr ){}\,{\sstt } -{2}\,{\bb}\,{}\Biggl ({5}\,{\Lx}-{3}\Biggr
){}\,{\utss}+{12}\,{\bb}\,{}\Biggl ({}-{2}\,{\Lx}+{\Ly}\Biggr ){}\,{\tttu } +{12}\,{\bb}\,{\Lx}\,{\tfiveou } \Biggr \}
\nonumber \\ &&
+ \Biggl \{ u \leftrightarrow t \Biggr \} \nonumber \\  
\MX{2}{+++-}&=&-\MP{2}{+++-}\nonumber \\
\MV{2}{+++-}&=&{ 0}\\
\MS{2}{++--}&=&{ }
-{4}\,{\bb}\,{\Lx^2}\,{\tfiveou } 
-{8}\,{\bb}\,{\Lx}\,{}\Biggl ({\Lx}+{1}\Biggr ){}\,{\sstt } 
\nonumber \\ &&
+{}\Biggl ({16}\,{\bb}\,{\Lidx}-{8}\,{\bb}\,{}\Biggl ({\Ly}-{2}\Biggr ){}\,{\Licy}-{8}\,{\bb}\,{\Licx}\,{\Ly}-{8}\,{\bb}\,{}\Biggl ({\Lx}-{\Ly}+{\Lx}\,{\Ly}\Biggr ){}\,{\Libx}
\nonumber \\ &&
-{16}\,{\bb}\,{\zeta_3}-{31\over 45}\,{\bb}\,{\pi^4}+{4\over 3}\,{\bb}\,{}\Biggl ({}-{3}\,{\Ly}-{3}-{\Lx^2}+{2}\,{\Lx}\,{\Ly}\Biggr ){}\,{\pi^2}
\nonumber \\ &&
+{1\over 3}\,{\bb}\,{}\Biggl ({}-{24}\,{\Lx^2}-{3}-{18}\,{\Lx^2}\,{\Ly^2}-{48}\,{\Lx}+{8}\,{\Ly^3}+{24}\,{\Lx}\,{\Ly}\Biggr ){}\Biggr ){}\,{\utss}
\nonumber \\ &&
+{}\Biggl ({}-{16}\,{\bb}\,{\Lidz}-{32}\,{\bb}\,{\Lidy}-{32}\,{\bb}\,{\Lidx}
\nonumber \\ &&
+{8}\,{\bb}\,{}\Biggl ({}-{2}+{\Ly}+{3}\,{\Lx}\Biggr ){}\,{\Licy}+{8}\,{\bb}\,{}\Biggl ({\Ly}+{3}\,{\Lx}+{2}\Biggr ){}\,{\Licx}+{8}\,{\bb}\,{\Liby}\,{\Lx}\,{\Ly}
\nonumber \\ &&
+{}\Biggl ({}-{8\over 3}\,{\bb}\,{\pi^2}+{8}\,{\bb}\,{}\Biggl ({}-{2}\,{\Ly}-{2}\,{\Lx}+{\Lx}\,{\Ly}\Biggr ){}\Biggr ){}\,{\Libx}+{112\over 45}\,{\bb}\,{\pi^4}
\nonumber \\ &&
+{4\over 3}\,{\bb}\,{}\Biggl ({}-{6}\,{\Lx}\,{\Ly}+{4}\,{\Ly}+{\Ly^2}+{3}-{2}\,{\Lx}+{\Lx^2}\Biggr ){}\,{\pi^2}
\nonumber \\ &&
-{4\over 3}\,{\bb}\,{}\Biggl ({}-{9}\,{\Lx^2}+{\Lx^4}+{6}\,{\Lx^2}\,{\Ly}-{9}\,{\Lx^2}\,{\Ly^2}-{4}\,{\Lx^3}\,{\Ly}+{6}\,{\Lx}\,{\Ly^2}-{6}\,{\Ly}-{2}\,{\Lx^3}+{2}\,{\Ly^3}
\nonumber \\ &&
+{6}\,{\Lx}+{6}\,{\Lx}\,{\Ly}\Biggr ){}\Biggr ){}\,{\tttu } 
\nonumber \\ && 
+{i\pi }\Biggl\{{ }-{8}\,{\bb}\,{}\Biggl ({1}+{2}\,{\Lx}\Biggr ){}\,{\sstt } 
-{8}\,{\bb}\,{\Lx}\,{\tfiveou } 
\nonumber \\ &&
+{}\Biggl ({}-{16}\,{\bb}\,{\Licy}-{8}\,{\bb}\,{\Lx}\,{\Liby}-{8}\,{\bb}\,{\Libx}\,{\Lx}+{8\over 3}\,{\bb}\,{\pi^2}\,{\Ly}
\nonumber \\ &&
+{4\over 3}\,{\bb}\,{}\Biggl ({6}\,{\Ly^2}-{12}-{6}\,{\Lx}\,{\Ly}-{9}\,{\Lx}\,{\Ly^2}+{\Ly^3}\Biggr ){}\Biggr ){}\,{\utss}
\nonumber \\ &&
+{}\Biggl ({32}\,{\bb}\,{\Licy}+{32}\,{\bb}\,{\Licx}+{32}\,{\bb}\,{\Ly}\,{\Liby}+{32}\,{\bb}\,{}\Biggl ({\Ly}-{1}\Biggr ){}\,{\Libx}
\nonumber \\ &&
-{4\over 3}\,{\bb}\,{}\Biggl ({}-{2}+{5}\,{\Ly}+{3}\,{\Lx}\Biggr ){}\,{\pi^2}
\nonumber \\ &&
-{4\over 3}\,{\bb}\,{}\Biggl
({}-{9}\,{\Lx^2}\,{\Ly}+{6}\,{\Ly^2}-{27}\,{\Lx}\,{\Ly^2}+{12}\,{\Lx}\,{\Ly}+{6}\,{\Ly}-{12}\,{\Lx}-{6}\,{\Lx^2}+{\Ly^3}+{3}\,{\Lx^3}\Biggr
){}\Biggr ){}\,{\tttu } 
\Biggr \}
\nonumber \\ &&
+ \Biggl \{ u \leftrightarrow t \Biggr \} \nonumber \\  
\label{eq:fermion}
\MF{2}{++--}&=&{ }-{2}\,{\bb}\,{\Lx^2}\,{\tfiveou } 
\nonumber \\ &&
+{}\Biggl ({16}\,{\bb}\,{\Licy}+{8}\,{\bb}\,{}\Biggl ({\Ly}-{\Lx}\Biggr ){}\,{\Libx}-{16}\,{\bb}\,{\zeta_3}-{2\over 3}\,{\bb}\,{}\Biggl ({5}+{6}\,{\Ly}\Biggr ){}\,{\pi^2}
\nonumber \\ &&
+{2\over 3}\,{\bb}\,{}\Biggl ({}-{6}\,{\Ly^2}+{12}\,{\Lx}\,{\Ly}+{4}\,{\Ly^3}-{3}-{12}\,{\Ly}\Biggr ){}\Biggr ){}\,{\utss}-{4}\,{\bb}\,{\Lx}\,{\sstt } 
\nonumber \\ &&
+{}\Biggl ({}-{16}\,{\bb}\,{\Lidz}-{16}\,{\bb}\,{\Lidy}-{16}\,{\bb}\,{\Lidx}+{8}\,{\bb}\,{}\Biggl ({}-{1}+{2}\,{\Lx}\Biggr ){}\,{\Licy}
\nonumber \\ &&
+{8}\,{\bb}\,{}\Biggl ({1}+{2}\,{\Lx}\Biggr ){}\,{\Licx}+{}\Biggl ({}-{8\over 3}\,{\bb}\,{\pi^2}-{8}\,{\bb}\,{}\Biggl ({\Lx}+{\Ly}\Biggr ){}\Biggr ){}\,{\Libx}+{7\over 9}\,{\bb}\,{\pi^4}
\nonumber \\ &&
-{2\over 3}\,{\bb}\,{}\Biggl ({\Ly^2}+{2}\,{\Lx}\,{\Ly}-{\Ly}-{\Lx}-{3}+{\Lx^2}\Biggr ){}\,{\pi^2}
-{1\over 3}\,{\bb}\,{}\Biggl ({5}\,{\Lx^4}-{6}\,{\Lx}\,{\Ly^2}-{12}\,{\Ly}-{10}\,{\Lx^3}+{\Ly^4}
\nonumber \\ &&
-{18}\,{\Lx^2}-{4}\,{\Lx}\,{\Ly^3}-{20}\,{\Lx^3}\,{\Ly}+{6}\,{\Lx^2}\,{\Ly^2}+{12}\,{\Lx}\,{\Ly}+{12}\,{\Lx}+{10}\,{\Ly^3}+{30}\,{\Lx^2}\,{\Ly}\Biggr ){}\Biggr ){}\,{\tttu } 
\nonumber \\ &&
+{i\pi }\Biggl\{{ }-{4}\,{\bb}\,{\sstt }+{}\Biggl ({4\over 3}\,{\bb}\,{\pi^2}+{8}\,{\bb}\,{}\Biggl ({}-{\Lx}\,{\Ly}+{\Ly}-{1}+{\Ly^2}\Biggr ){}\Biggr ){}\,{\utss}
-{4}\,{\bb}\,{\Lx}\,{\tfiveou } 
\nonumber \\ &&
+{}\Biggl ({16}\,{\bb}\,{\Licy}+{16}\,{\bb}\,{\Licx}+{16}\,{\bb}\,{\Ly}\,{\Liby}+{16}\,{\bb}\,{}\Biggl ({\Ly}-{1}\Biggr ){}\,{\Libx}
\nonumber \\ &&
-{4\over 3}\,{\bb}\,{}\Biggl ({}-{1}+{2}\,{\Ly}+{2}\,{\Lx}\Biggr ){}\,{\pi^2}
\nonumber \\ &&
+{4\over 3}\,{\bb}\,{}\Biggl ({12}\,{\Lx}\,{\Ly^2}-{3}\,{\Ly}+{6}\,{\Lx^2}\,{\Ly}-{3}\,{\Ly^2}-{2}\,{\Lx^3}+{6}\,{\Lx}+{3}\,{\Lx^2}-{6}\,{\Lx}\,{\Ly}\Biggr ){}\Biggr ){}\,{\tttu } 
\Biggr \}
\nonumber \\ &&
+ \Biggl \{ u \leftrightarrow t \Biggr \}  \\  
\MP{2}{++--}&=&{ 6}\,{\bb}\,{\Lx^2}\,{\tfiveou } 
+{4}\,{\bb}\,{\Lx}\,{}\Biggl ({4}\,{\Lx}+{3}\Biggr ){}\,{\sstt } 
\nonumber \\ &&
+{}\Biggl ({}-{32}\,{\bb}\,{\Lidx}+{16}\,{\bb}\,{\Licy}\,{\Ly}+{16}\,{\bb}\,{}\Biggl ({\Ly}-{1}\Biggr ){}\,{\Licx}+{8}\,{\bb}\,{}\Biggl ({\Ly}-{\Lx}\Biggr ){}\,{\Liby}
\nonumber \\ &&
+{16}\,{\bb}\,{\Libx}\,{\Ly}\,{\Lx}+{16}\,{\bb}\,{\zeta_3}+{47\over 45}\,{\bb}\,{\pi^4}-{2\over 3}\,{\bb}\,{}\Biggl ({6}\,{\Lx}\,{\Ly}-{2}\,{\Lx^2}-{6}\,{\Lx}-{5}\Biggr ){}\,{\pi^2}
\nonumber \\ &&
-{2\over 3}\,{\bb}\,{}\Biggl ({}-{15}\,{\Lx^2}\,{\Ly^2}-{18}\,{\Ly^2}+{12}\,{\Lx}\,{\Ly}-{6}-{36}\,{\Ly}-{4}\,{\Lx^3}\,{\Ly}+{\Ly^4}+{4}\,{\Ly^3}\Biggr ){}\Biggr ){}\,{\utss}
\nonumber \\ &&
+{}\Biggl ({16}\,{\bb}\,{\Lidz}+{48}\,{\bb}\,{\Lidy}+{48}\,{\bb}\,{\Lidx}-{8}\,{\bb}\,{}\Biggl ({}-{3}+{2}\,{\Ly}+{4}\,{\Lx}\Biggr ){}\,{\Licy}
\nonumber \\ &&
-{8}\,{\bb}\,{}\Biggl ({2}\,{\Ly}+{4}\,{\Lx}+{3}\Biggr ){}\,{\Licx}-{16}\,{\bb}\,{\Liby}\,{\Lx}\,{\Ly}
\nonumber \\ &&
+{}\Biggl ({8\over 3}\,{\bb}\,{\pi^2}-{8}\,{\bb}\,{}\Biggl ({}-{3}\,{\Ly}+{2}\,{\Lx}\,{\Ly}-{3}\,{\Lx}\Biggr ){}\Biggr ){}\,{\Libx}-{16\over 5}\,{\bb}\,{\pi^4}
\nonumber \\ &&
-{2\over 3}\,{\bb}\,{}\Biggl ({11}\,{\Ly}+{2}\,{\Ly^2}
-{16}\,{\Lx}\,{\Ly}-{5}\,{\Lx}+{9}+{2}\,{\Lx^2}\Biggr ){}\,{\pi^2}
\nonumber \\ &&
+{2\over 3}\,{\bb}\,{}\Biggl ({}-{18}\,{\Ly}+{3}\,{\Lx^4}+{15}\,{\Lx}\,{\Ly^2}-{27}\,{\Lx^2}-{7}\,{\Lx^3}+{\Ly^4}-{30}\,{\Lx^2}\,{\Ly^2}-{4}\,{\Lx}\,{\Ly^3}
\nonumber \\ &&
-{12}\,{\Lx^3}\,{\Ly}+{18}\,{\Lx}\,{\Ly}+{18}\,{\Lx}+{7}\,{\Ly^3}+{21}\,{\Lx^2}\,{\Ly}\Biggr ){}\Biggr ){}\,{\tttu } 
\nonumber \\ &&
+{i\pi }\Biggl\{{ 4}\,{\bb}\,{}\Biggl ({3}+{8}\,{\Lx}\Biggr ){}\,{\sstt } 
+{12}\,{\bb}\,{\Lx}\,{\tfiveou } 
\nonumber \\ &&
+{}\Biggl ({32}\,{\bb}\,{\Licx}+{16}\,{\bb}\,{\Lx}\,{\Liby}+{16}\,{\bb}\,{\Libx}\,{\Lx}-{4\over 3}\,{\bb}\,{}\Biggl ({4}\,{\Ly}+{1}\Biggr ){}\,{\pi^2}
\nonumber \\ &&
-{8\over 3}\,{\bb}\,{}\Biggl ({}-{3}\,{\Ly}-{9}\,{\Lx}\,{\Ly^2}-{3}\,{\Lx}\,{\Ly}+{\Ly^3}+{3}\,{\Lx^2}-{9}\Biggr ){}\Biggr ){}\,{\utss}
\nonumber \\ &&
+{}\Biggl ({}-{48}\,{\bb}\,{\Licy}-{48}\,{\bb}\,{\Licx}-{48}\,{\bb}\,{\Ly}\,{\Liby}-{48}\,{\bb}\,{}\Biggl ({\Ly}-{1}\Biggr ){}\,{\Libx}
\nonumber \\ &&
+{4\over 3}\,{\bb}\,{}\Biggl ({}-{3}+{8}\,{\Ly}+{4}\,{\Lx}\Biggr ){}\,{\pi^2}+{4\over 3}\,{\bb}\,{}\Biggl ({2}\,{\Ly^3}+{9}\,{\Ly}+{9}\,{\Ly^2}-{18}\,{\Lx}
\nonumber \\ &&
-{12}\,{\Lx^2}\,{\Ly}-{42}\,{\Lx}\,{\Ly^2}
+{4}\,{\Lx^3}-{9}\,{\Lx^2}+{18}\,{\Lx}\,{\Ly}\Biggr ){}\Biggr ){}\,{\tttu } 
\Biggr \}
\nonumber \\ &&
+ \Biggl \{ u \leftrightarrow t \Biggr \} \nonumber \\  
\MX{2}{++--}&=& 
-{6}\,{\bb}\,{\Lx^2}\,{\tfiveou }  
-{12}\,{\bb}\,{\Lx}\,{}\Biggl ({2}\,{\Lx}+{1}\Biggr ){}\,{\sstt } \nonumber \\ &&
+{}\Biggl ({32}\,{\bb}\,{\Lidy}-{16}\,{\bb}\,{\Lx}\,{\Licx}-{16}\,{\bb}\,{\Lx}\,{\Licy}-{47\over 45}\,{\bb}\,{\pi^4}-{2\over
3}\,{\bb}\,{}\Biggl ({}-{1}+{2}\,{\Lx^2}\Biggr ){}\,{\pi^2} \nonumber \\ &&
+{2\over 3}\,{\bb}\,{}\Biggl
({}-{3}-{4}\,{\Lx^3}\,{\Ly}-{18}\,{\Ly^2}-{3}\,{\Lx^2}\,{\Ly^2}+{\Ly^4}-{36}\,{\Ly}\Biggr ){}\Biggr ){}\,{\utss}\nonumber \\ &&
+{}\Biggl ({}-{48}\,{\bb}\,{\Lidy}-{48}\,{\bb}\,{\Lidx}+{24}\,{\bb}\,{}\Biggl ({1}+{\Ly}+{\Lx}\Biggr ){}\,{\Licx}+{24}\,{\bb}\,{}\Biggl ({\Ly}+{\Lx}-{1}\Biggr ){}\,{\Licy} \nonumber \\ &&
-{24}\,{\bb}\,{}\Biggl ({\Ly}+{\Lx}\Biggr ){}\,{\Libx}+{47\over 15}\,{\bb}\,{\pi^4}+{2\over 3}\,{\bb}\,{}\Biggl ({13}\,{\Ly}+{3}\,{\Lx^2}+{9}-{7}\,{\Lx}+{3}\,{\Ly^2}-{12}\,{\Lx}\,{\Ly}\Biggr ){}\,{\pi^2} \nonumber \\ &&
-{1\over 3}\,{\bb}\,{}\Biggl ({}-{12}\,{\Lx^3}\,{\Ly}-{36}\,{\Ly}+{36}\,{\Lx}-{10}\,{\Lx^3}+{3}\,{\Lx^4}-{54}\,{\Lx^2}+{30}\,{\Lx^2}\,{\Ly}+{3}\,{\Ly^4} \nonumber \\ &&
+{10}\,{\Ly^3}-{12}\,{\Lx}\,{\Ly^3}+{42}\,{\Lx}\,{\Ly^2}+{36}\,{\Lx}\,{\Ly}-{18}\,{\Lx^2}\,{\Ly^2}\Biggr ){}\Biggr ){}\,{\tttu } \nonumber \\ &&
+{i\pi }\Biggl\{
-{12}\,{\bb}\,{}\Biggl ({1}+{4}\,{\Lx}\Biggr ){}\,{\sstt }  
-{12}\,{\bb}\,{\Lx}\,{\tfiveou } \nonumber \\ &&
+{}\Biggl ({}-{32}\,{\bb}\,{\Licy}+{8\over 3}\,{\bb}\,{\Lx}\,{\pi^2}+{8\over 3}\,{\bb}\,{}\Biggl
({}-{9}+{\Ly^3}-{3}\,{\Lx^2}\,{\Ly}-{9}\,{\Ly}\Biggr ){}\Biggr ){}\,{\utss}\nonumber \\ &&
+{}\Biggl ({48}\,{\bb}\,{\Licx}+{48}\,{\bb}\,{\Licy}-{48}\,{\bb}\,{\Libx}-{4}\,{\bb}\,{}\Biggl ({\Ly}+{\Lx}-{1}\Biggr ){}\,{\pi^2}\nonumber \\ &&
-{4}\,{\bb}\,{}\Biggl
({6}\,{\Lx}\,{\Ly}+{\Lx^3}-{3}\,{\Lx^2}-{3}\,{\Lx}\,{\Ly^2}-{6}\,{\Lx}-{3}\,{\Lx^2}\,{\Ly}+{3}\,{\Ly^2}+{3}\,{\Ly}+{\Ly^3}\Biggr
){}\Biggr ){}\,{\tttu } \Biggr\}
\nonumber \\ &&
+ \Biggl \{ u \leftrightarrow t \Biggr \} \nonumber \\  
\MV{2}{++--}&=&{ }-1
\end{eqnarray}
The contribution from fermion exchange $\MF{2}{}$ has previously 
been computed in
Ref.~\cite{Bern:2001dg} and we find complete 
agreement with the results presented
there once the different definitions of the helicity amplitudes are taken into
account.
\end{document}